\documentclass[review,authoryear,12pt]{elsarticle}

\usepackage[rgb, x11names]{xcolor} 
\usepackage{amsfonts, amsmath, amssymb, bm} 
\usepackage{hyperref} 
\usepackage{paralist} 
\usepackage{breqn} 
\usepackage{marginnote} 
\usepackage{todonotes} 
\usepackage{booktabs} 
\usepackage{array} 
\usepackage{graphicx} 
\usepackage{marvosym} 
\usepackage{bookmark}
\usepackage[tight]{subfigure}
\usepackage{multirow}
\usepackage{hhline}

\usepackage{tikz}
\usepackage{pgfplotstable}
\usepackage{pgfplots}
\usepackage{tikz-3dplot}
\usetikzlibrary{mindmap, trees}
\usetikzlibrary{arrows,decorations.pathmorphing,decorations.markings,backgrounds,positioning,fit,plotmarks,shadows,intersections,patterns}
\pgfdeclarelayer{background}
\pgfdeclarelayer{foreground}
\pgfsetlayers{background,main,foreground}
\pgfplotsset{compat=1.3,every axis/.append style={font=\scriptsize}, every legend/.append style={font=\scriptsize}}
\tikzstyle{mylabel} = [text=orange, ultra thick, inner sep=1pt, minimum size=15pt, yshift=-6pt, xshift=30pt]
\tikzstyle{fancylabel} = [rounded corners, fill=Blue4, draw=black, very thick, text=white, inner sep=0pt, minimum size=15pt, yshift=0pt]
\tikzstyle{labelwithbackground} = [text=red, fill=none, ultra thick, inner sep=1pt, minimum size=12pt, yshift=-9pt, xshift=9pt]

\newlength{\colorbarlength}

\newlength{\defbaselineskip}
\DeclareMathOperator{\argmin}{arg\,min}

\newcommand\blfootnote[1] 
{
	\begingroup
	
	\footnotetext{#1}
	\addtocounter{footnote}{-1}
	\endgroup
}
\usepackage{environ}

\label{packages}

\journal{}
\begin{document}
\begin{frontmatter}

\title{\textbf{Mitigating Gyral Bias in Cortical Tractography via Asymmetric Fiber Orientation Distributions}}

\author[unc]{Ye Wu}
\author[unc]{Yoonmi Hong}
\author[zjut]{Yuanjing Feng\corref{scor}}
\ead{fyjing@zjut.edu.cn}
\author[unc,ku]{Dinggang Shen\corref{scor}}
\ead{dgshen@med.unc.edu}
\author[unc]{Pew-Thian~Yap\corref{pcor}}
\ead{ptyap@med.unc.edu}
\cortext[pcor]{Primary corresponding author}
\cortext[scor]{Secondary corresponding authors}

\address[unc]{Department of Radiology and Biomedical Research Imaging Center (BRIC)\\University of North Carolina at Chapel Hill, NC, U.S.A.}
\address[zjut]{Institute of Information Processing and Automation, Zhejiang University of Technology, Hangzhou, China}
\address[ku]{Department of Brain and Cognitive Engineering, Korea University, Seoul, Republic of Korea}

\begin{abstract}
Diffusion tractography in brain connectomics often involves tracing axonal trajectories across gray-white matter boundaries in gyral blades of complex cortical convolutions.
To date, gyral bias is observed in most tractography algorithms with streamlines predominantly terminating at gyral crowns instead of sulcal banks.
This work demonstrates that asymmetric fiber orientation distribution functions (AFODFs), computed via a multi-tissue global estimation framework, can mitigate the effects of gyral bias, enabling fiber streamlines at gyral blades to make sharper turns into the cortical gray matter.
We use ex-vivo data of an adult rhesus macaque and in-vivo data from the Human Connectome Project (HCP) to show that the fiber streamlines given by AFODFs bend more naturally into the cortex than the conventional symmetric FODFs in typical gyral blades.
We demonstrate that AFODF tractography improves cortico-cortical connectivity and provides highly consistent outcomes between two different field strengths (3T and 7T).
\end{abstract}


\end{frontmatter}

\section{Introduction}
\label{sec:Introduction}

Diffusion magnetic resonance imaging (DMRI) \citep{johansen2013diffusion} is a powerful imaging technique for non-invasive quantification of tissue microstructure and mapping of axonal trajectories by probing the diffusion patterns of water molecules in the living human brain.
DMRI-based connectomics is widely used for understanding neurological development associated with the cerebral cortex \citep{johansen2013diffusion,yamada2009mr,mori2002fiber}.

A white matter (WM) fiber tract is a collection of axons in the central nervous system with common origin and destination sites, typically in the cortical gray matter (GM) \citep{makris1997morphometry,noback2005human,robertson2016neuroanatomical}.
These tracts have long range and involve axons transversing from cortical GM sites through the superficial WM, then the deep WM, and into a distant cortical or subcortical structure. 
For an accurate connectivity map of the brain, estimated fiber streamlines must be able not only to follow major fiber bundles through the deep white matter, but must also correctly follow axonal trajectories as they cross the white matter/gray matter (WM-GM) boundary. 
However, it is shown that up to $70\%$ of the streamlines produced by the state-of-the-art tractography algorithms actually do not reach the GM, even with high angular resolution diffusion imaging (HARDI) \citep{cote2013tractometer,Maier-Hein2017}.
This can be attributed to the complexity of fiber arrangement in the superficial WM, which resides beneath the cortical sheet \citep{reveley2015superficial}.
This in turn causes gyral bias, with streamlines preferentially terminating at gyral crowns rather than sulcal banks  \citep{van2014mapping}.
Gyral bias stems from the technical difficulties in tracing highly-curved axonal trajectories across WM-GM boundaries in gyral blades \citep{van2014mapping}.
The shortcomings of existing tractography algorithms may lead to severe bias in connectivity analysis \citep{reveley2015superficial}.

Gyral bias can be mitigated by increasing the spatial resolution of DMRI data \citep{sotiropoulos2016fusion,heidemann2012k} or imposing constraints derived from anatomical (e.g., T1-weighted) images \citep{St-Onge2018,teillac2017novel,smith2012anatomically}.  In this paper, we will however introduce an approach to mitigate the gyral bias that does not rely on time-consuming and expensive high-resolution data as well as hypothetical cortical connections derived from anatomical images. 
Our technique utilizes subvoxel asymmetry and fiber continuity to improve cortical tractography of highly-curved axonal trajectories. 


The orientation information at each voxel is typically encoded as a fiber orientation distribution function (FODF), which is typically computed by using spherical deconvolution (SD) \citep{tournier2004direct,Tournier2007}.
The SD method assumes that the signal profile can be represented as the convolution of the FODF with a fiber response function estimated from voxels that contain coherent single-directional axonal bundles.
FODFs are typically assumed to be antipodal symmetric, implying that an orientation in a positive hemisphere is always identical to its counterpart in the negative hemisphere. 
However, this symmetry limits the FODF in representing complex WM geometries such as fanning and bending.

The basic idea of this paper is to incorporate information from neighboring voxels in estimating \emph{asymmetric} FODFs (AFODFs). To date, the impact of AFODFs on the gyral bias has not been investigated.
We extend the multi-tissue model in \cite{Jeurissen2014} to account for the different polarity of fiber orientations separately in order to capture the asymmetry of the underlying fiber geometry in a local neighborhood. We estimate the AFODF at each voxel by enforcing orientation continuity across voxels. We assume that a fiber streamline leaves a voxel along direction $\mathbf{u}$ and enters a neighboring voxel along the reverse direction $\mathbf{-u}$ with higher fiber continuity. Our continuity constraint is constructed to minimize the difference between the two fiber orientation within the neighborhood.

Unlike \cite{karayumak2018asymmetric,bastiani2017improved,reisert2012geometry}, our approach is formulated as a convex problem and does not require initialization with pre-computed symmetric FODFs. Instead, we estimate the AFODFs directly from the data by imposing the fiber continuity constraint across voxels.
Moreover, our algorithm provides the global solution for multiple voxels simultaneously in adherence to the fiber continuity constraint instead of solving for each voxel individually. The convexity of our problem formulation allows efficient optimization for large-scale global solutions, unlike the non-convex formulation in \cite{auria2015structured}.

\begin{figure*}[!t]\centering
		\includegraphics[width=\textwidth]{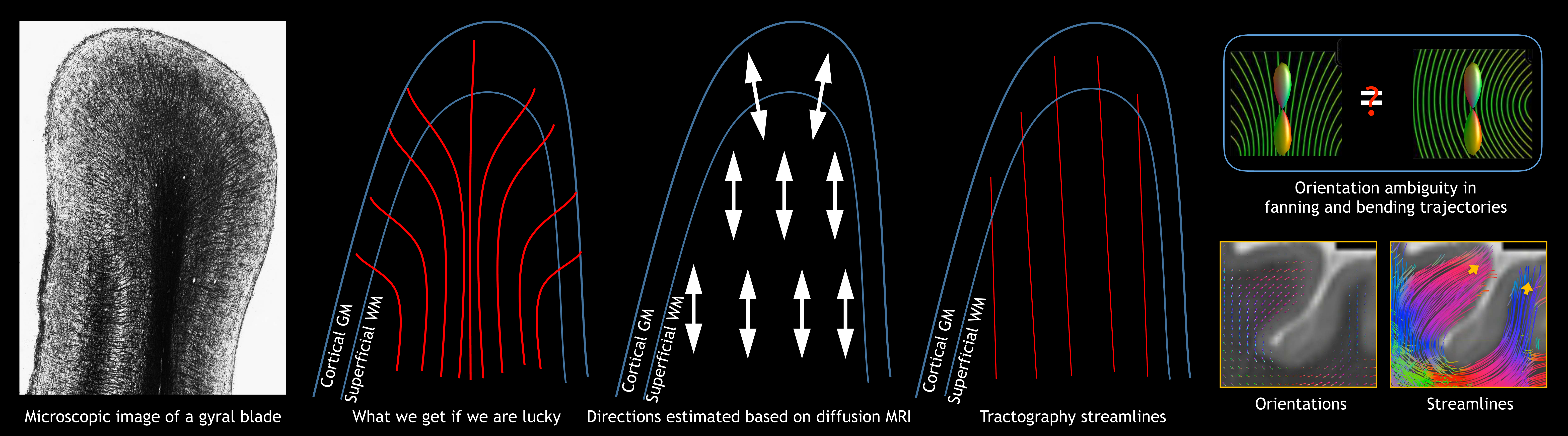}\\		
	\caption{Existing tractography algorithms typically produce biased fiber streamlines that predominantly terminate at gyral crowns, which contradicts actual histology showing that axonal trajectories in gyral blades traverse perpendicularly into gyral crowns and sulcal banks. Gyral bias is caused by the ambiguity of the fiber orientation distribution functions (FODFs) in voxels traversed by bending and fanning trajectories. 
    The inability to trace the fiber streamlines correctly in the superficial WM impedes the detection of long-range cortical connections \citep{reveley2015superficial}.}\vspace{-10pt}
	\label{fig:complex}
\end{figure*}

Part of this work was presented in \cite{ye2018afodf}. Herein, we
\begin{inparaenum}[(i)]
	\item provide a more detailed description of the proposed method;
	\item devise a novel tractography algorithm;
	\item reformulate the fiber continuity term using the von Mises-Fisher distribution;
	\item optimize the algorithm for accelerated global solutions of multiple voxels;
	\item provide new insights into consistency across scanners and spatial resolutions; and
	\item compare fiber densities at gyral crowns and sulcal banks determined based on histology and diffusion tractography.
\end{inparaenum}
None of these is part of the conference publication.

We describe our method in detail in the \nameref{sec:Method} section and demonstrate its effectiveness with in-vivo data in the \nameref{sec:Experiments} section. We  provide further discussion in the \nameref{sec:Discussion} section before concluding in the \nameref{sec:Conclusion} section.

\section{Methods}\label{sec:Method}

\subsection {Multi-Tissue Asymmetric Fiber Orientation Distribution}\label{sec:AFODFs}

The FODF $F(\mathbf{u})$, $\mathbf{u} \in \mathbb{S}^2$ usually assumes antipodal symmetry in the sense that $F(\mathbf{u})=F(-\mathbf{u})$.
However, this symmetry limits the FODF in representing complex configurations such as fanning and bending, both very common in gyral blades (see Figure~\ref{fig:complex}). 
We extend the multi-tissue constrained spherical deconvolution (MT-CSD) framework \citep{Jeurissen2014} by incorporating information from voxel neighborhood $\mathcal{N}$ to allow FODF asymmetry for better representation of complex configurations.


MT-CSD decomposes the diffusion signal $S_{\mathbf{p}}(\mathbf{g})$, for gradient direction $\mathbf{g}\in\mathbb{S}^2$ and location $\mathbf{p} \in \mathbb{R}^3$, into $M \in \mathbb{N}^{+}$ tissue types, each of which is represented by the spherical convolution of an axially symmetric response function (RF) $R_i(\mathbf{g},\cdot)$ \citep{Tournier2007,Jeurissen2014} with an FODF $F_{\mathbf{p},i}(\cdot)$, $1\leq i\leq M$, giving an overall signal
\begin{equation}
	S_{\mathbf{p}}(\mathbf{g}) = \int_{\mathbf{u}\in\mathbb{S}^2} \sum_{i=1}^{M} R_i (\mathbf{g},\mathbf{u})F_{\mathbf{p},i}(\mathbf{u})d\mathbf{u}.
\end{equation}
A typical multi-tissue model accounts for signal contributions from white matter (WM), gray matter (GM),
and cerebrospinal fluid (CSF), with $M=3$. Based on the observation that the GM and CSF compartments are isotropic, the asymmetry will be considered only for the WM compartment. We drop the index $i$ for simplicity.

To improve robustness to noise, FODFs are often estimated with some form of regularization  \citep{Tournier2007,auria2015structured}.
We regularize the estimation of AFODFs via the fiber continuity constraint. A fiber streamline leaving a voxel $\mathbf{p}$ along direction $\mathbf{u}$ should enter the adjacent voxel $\mathbf{q}\in\mathcal{N}_{\mathbf{p}}$ along direction $-\mathbf{u}$ with higher possibility. Here, we denote the $\mathcal{N}_{\mathbf{p}}$ as a second order neighbour in the center $\mathbf{p}$.   We measure the discontinuity $\Phi(\cdot)$ of the FODFs of $N$ voxels, denoted as $\mathbf{F}$, over direction $\mathbf{u}\in\mathbb{S}^2$ by
\begin{equation}
\Phi(\mathbf{F})=
\sum_{\mathbf{p}}\int_{\mathbf{u}\in\mathbb{S}^2}
\left| F_{\mathbf{p}}(\mathbf{u}) - \frac{1}{K_{\mathbf{p},\mathbf{u}}}\sum_{\mathbf{q} \in \mathcal{N}_{\mathbf{p}}
} W(\langle\mathbf{\hat{v}_{p,q}}, \mathbf{u}\rangle) F_{\mathbf{q}}(\mathbf{u})\right|d\mathbf{u},\label{eq:discon}
\end{equation}
where $W(\langle\mathbf{\hat{v}_{p,q}}, \mathbf{u}\rangle)$ denotes a directional probability distribution function (PDF) that is decided by the angular difference between $\mathbf{u}$ and $\mathbf{\hat{v}_{p,q}} = \frac{\mathbf{q} - \mathbf{p}}{\| \mathbf{q} - \mathbf{p} \|}$. $K_{\mathbf{p},\mathbf{u}} = \sum_{\mathbf{q} \in \mathcal{N}_{\mathbf{p}}
}
W(\langle\mathbf{\hat{v}_{p,q}}, \mathbf{u}\rangle)$ is a normalization term. We choose $W(\langle\mathbf{\hat{v}_{p,q}}, \cdot \rangle)$ to be the von Mises-Fisher distribution with reference direction $\mathbf{\hat{v}_{p,q}}$ (see \nameref{sec:Prior}).

FODF is commonly represented as a series of real spherical harmonics (SHs) with even-order \citep{tournier2004direct,Frank2002}.
Odd-order SHs typically capture only noise \citep{hess2006q} and are therefore not included in the representation.
Unlike the method in \cite{bastiani2017improved}, which attempts to capture the asymmetry using an additional component represented by odd-order SHs, we represent the two fiber orientations with different polarity as $\mathbb{S}_{+}^2$ and $\mathbb{S}_{-}^2$ of the WM AFODF separately using
\begin{equation}\label{eq:afodf_sh}
	F_{\mathbf{p}}(\mathbf{u})=
	\begin{bmatrix}
		\mathbf{Y}^{+}(\mathbf{u}) & 0\\
		0 & \mathbf{Y}^{-}(\mathbf{u})
	\end{bmatrix}
	\begin{bmatrix}
		\mathbf{X}_{\text{WM}}^{+}(\mathbf{p})\\
		\mathbf{X}_{\text{WM}}^{-}(\mathbf{p})\\
	\end{bmatrix},
\end{equation}
where $\mathbf{Y}^{+}(\mathbf{u}), \,\mathbf{u}\in\mathbb{S}_{+}^2$, and $\mathbf{Y}^{-}(\mathbf{u}), \,\mathbf{u}\in\mathbb{S}_{-}^2$, are the real symmetric SH bases that are defined on different hemispheres of $\mathbb{S}^2$. $\mathbf{X}_{\text{WM}}^{+}(\mathbf{p})$ and $\mathbf{X}_{\text{WM}}^{-}(\mathbf{p})$ are the corresponding SH coefficients. 
Note that the fiber continuity constraint is applied on the entire FODF so that there is no discontinuity between the positive and negative hemispheres.
In this paper, we set the maximum SH order as 8. Following \citep{Jeurissen2014}, we use a multi-tissue formulation and represent the GM and CSF FODFs using zeroth order SH coefficients.

\subsection {Directional PDF}\label{sec:Prior}
The directional PDF should decrease monotonically with the angular difference between the direction $-\mathbf{u}$ and a reference direction $\mathbf{\hat{v}_{p,q}}$. We employ a von Mises-Fisher distribution over the sphere \citep{mardia2000distributions} and set
\begin{equation}
	W(\langle\mathbf{\hat{v}_{p,q}}, \mathbf{u}\rangle | \kappa)
	= \frac{\kappa^{1/2}}{(2\pi)^{3/2}I_{1/2}(\kappa)}\exp\left(-\kappa \langle \mathbf{\hat{v}_{p,q}}, \mathbf{u} \rangle\right),
\end{equation}
\noindent
where $\kappa \geq 0$ and $I_{1/2}$ denotes the modified Bessel function of the first kind at order $1/2$.
The greater the value of $\kappa$ the higher the concentration of the distribution around the reference direction $\mathbf{\hat{v}_{p,q}}$.
In particular, when $\kappa=0$, the distribution is uniform over the sphere, and as $\kappa \rightarrow \infty$, the distribution tends to a point density.
The distribution is rotationally symmetric around $\mathbf{\hat{v}_{p,q}}$ and is unimodal for $\kappa\ge 0$. In our experiment, $\kappa$ is set to 4.

\begin{figure*}[!t]\centering
		\includegraphics[width=\textwidth]{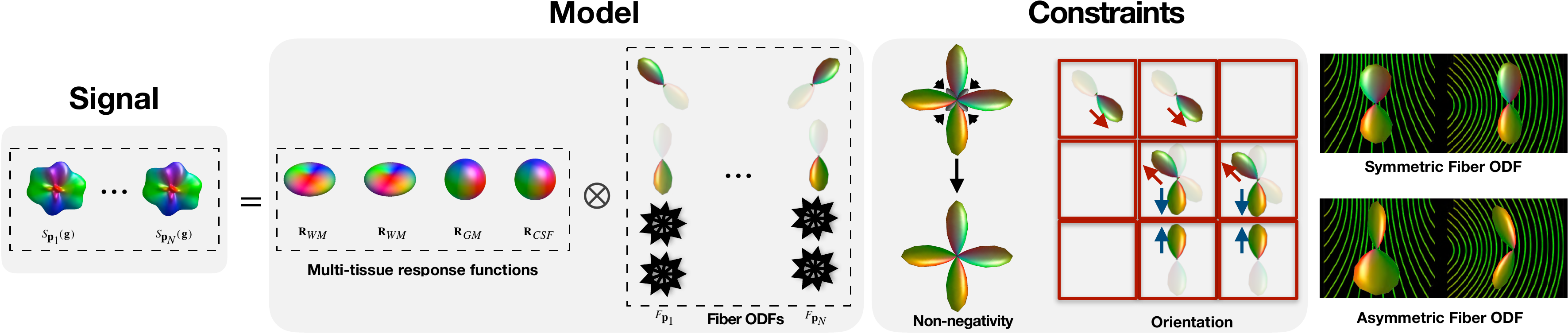}
	\caption{
To mitigate gyral bias, we allow the FODF to be asymmetric. This asymmetry is estimated with the orientation consistency constraint across voxels. This constraint links all voxels and necessitates their AFODFs to be estimated concurrently. The signal is represented as the convolution (`$\otimes$') of multi-tissue response functions with the asymmetric FODFs.
  }\vspace{-10pt}
	\label{fig:overview}
\end{figure*}

\subsection{Problem Formulation}\label{sec:CostFunction}
We group the signal vectors of $N$ voxels at locations $\mathbb{P}=\{\mathbf{p}_1, \mathbf{p}_2, \hdots, \mathbf{p}_N\}$ as columns of matrix $\mathbf{S}$ and the SH coefficients of the corresponding AFODFs as columns in matrix $\mathbf{X}$ (Figure~\ref{fig:overview}). With a set of directions $\mathbb{U} = \{\mathbf{u}_{1},\mathbf{u}_{2},\hdots,\mathbf{u}_{K}\}$, we aim to solve the SH coefficients of the $N$ voxels simultaneously as follows:
\begin{equation}
	\begin{aligned}
		\mathbf{\hat{X}} = \arg\min\limits_{\mathbf{X}} \left\| \mathbf{RX} - \mathbf{S} \right\|_F^2~\text{s.t.}~ \mathbf{AX} \succeq 0~ \text{and}~\Phi(\mathbf{F}) \leq \delta,\label{eq:model1}
	\end{aligned}
\end{equation}
where
\begin{equation*}
	\begin{aligned}
		\mathbf{X} =
		\begin{bmatrix}
			\mathbf{X}_{\text{WM}}^{+}(\mathbf{p}_1) & \hdots & \mathbf{X}_{\text{WM}}^{+}(\mathbf{p}_N)\\
			\mathbf{X}_{\text{WM}}^{-}(\mathbf{p}_1)  & \hdots  & \mathbf{X}_{\text{WM}}^{-}(\mathbf{p}_N)\\
			\mathbf{X}_{\text{GM}}(\mathbf{p}_1) & \hdots  & \mathbf{X}_{\text{GM}}(\mathbf{p}_N)\\
			\mathbf{X}_{\text{CSF}}(\mathbf{p}_1) & \hdots  & \mathbf{X}_{\text{CSF}}(\mathbf{p}_N)
		\end{bmatrix}
		\quad
	\end{aligned}
\end{equation*}
and $\mathbf{R} = \left[{\mathbf{R}_{\text{WM}}, \mathbf{R}_{\text{WM}}, \mathbf{R}_{\text{GM}}, \mathbf{R}_{\text{CSF}} }\right]$.
Matrix $\mathbf{R}$ maps the SH coefficients to the diffusion signal by spherical convolution.
Matrix $\mathbf{A}$ maps the SH coefficients to the AFODF amplitudes at directions $\mathbb{U}$ for imposing AFODF nonnegativity constraint by $\mathbf{AX} \succeq 0$ for all tissue types.
In \eqref{eq:model1}, $\delta$ controls the strictness of the fiber continuity constraint and $\|\cdot\|_F$ is the Frobenius norm.
$\Phi(\mathbf{F})$, defined in \eqref{eq:discon}, is computed only for the WM AFODF. We let
\begin{equation}
	\widehat{\mathbf{W}}(\mathbf{u})=
	\begin{bmatrix}
		\widehat{W}(\mathbf{p}_{1},\mathbf{p}_{1},\mathbf{u}) & \hdots & \widehat{W}(\mathbf{p}_{N},\mathbf{p}_{1},\mathbf{u})\\
		\vdots & \ddots & \vdots\\
		\widehat{W}(\mathbf{p}_{1},\mathbf{p}_{N},\mathbf{u}) & \hdots & \widehat{W}(\mathbf{p}_{N},\mathbf{p}_{N},\mathbf{u})\\
	\end{bmatrix},
\end{equation}
with
\begin{equation}
	\widehat{W}(\mathbf{p},\mathbf{q},\mathbf{u}) =
	\begin{cases}
		\frac{W(\langle\mathbf{\hat{v}_{p,q}}, \mathbf{u}\rangle)}{K_{\mathbf{p},\mathbf{u}}} & \mathbf{q} \in \mathcal{N}_{\mathbf{p}},\\
		0 & \text{otherwise},\\
	\end{cases}
\end{equation}
and the normalization term $K_{\mathbf{p},\mathbf{u}}$ defined in~\eqref{eq:discon}. We further define the operator $\circ$ as
\begin{equation}
	\mathbf{AX} \circ \widehat{\mathbf{W}} =
	\begin{bmatrix}
		( \mathbf{A}\mathbf{X} )_{1} \widehat{\mathbf{W}}(\mathbf{u}_{1}) \\
		\vdots \\
		( \mathbf{A}\mathbf{X} )_{K} \widehat{\mathbf{W}}(\mathbf{u}_{K})
	\end{bmatrix},
\end{equation}%
where $( \mathbf{A}\mathbf{X} )_{i}$ is the $i$-th row of $\mathbf{A}\mathbf{X}$ and corresponds to direction $\mathbf{u}_{i}$, $1\leq i \leq K$. Then, the problem~\eqref{eq:model1} can be rewritten as
\begin{equation}
	\begin{aligned}
		\mathbf{\hat{X}} = \arg\min\limits_{\mathbf{X}} \left\| \mathbf{RX} - \mathbf{S} \right\|_F^2 ~\text{s.t.}~ \mathbf{AX} \succeq 0 ~\text{and}~
		\left\| \mathbf{AX}\circ\widehat{\mathbf{W}} - \mathbf{AX} \right\|_F
		\leq \delta,\label{eq:model3}
	\end{aligned}
\end{equation}
which is a linear least-squares problem with linear inequality constraints that can be solved efficiently using convex optimization.

\subsection{Optimization}\label{sec:Optimization}
We solve the high-dimensional optimization problem~\eqref{eq:model3} using the alternating direction method of multipliers (ADMM) \citep{boyd2011distributed}.  We note that \eqref{eq:model3} can be equivalently written as
\begin{equation}
	\begin{aligned}
		\min\limits_{\mathbf{X},\mathbf{Z}} \frac{1}{2} \left\| \mathbf{RX} - \mathbf{S} \right\|_F^2 + \frac{\lambda}{2} \left\| \mathbf{AX} -\mathbf{AZ} \circ \widehat{\mathbf{W}}\right\|_F^2
		\\
		\text{s.t.}~  \mathbf{AX}  = \mathbf{AZ} \circ \widehat{\mathbf{W}}, ~\mathbf{AX}\succeq 0, ~\mathbf{AZ}\succeq 0,\label{eq:model4}
	\end{aligned}
\end{equation}
where $\mathbf{Z}$ is an auxiliary variable.
Let $\mathbf{H}$ be the multiplier for $\mathbf{AX}  - \mathbf{AZ} \circ \mathbf{\hat{W}}=\boldsymbol{0}$. Then, each iteration of ADMM involves minimizing the augmented Lagrangian
\begin{equation}
	\begin{aligned}
		L_{\lambda}(\mathbf{X},\mathbf{Z},\mathbf{H}) = \frac{1}{2} \left\| \mathbf{RX} - \mathbf{S} \right\|_F^2 +  I_{+}(\mathbf{AX})  \\ + \frac{\lambda}{2} \left\| \mathbf{AX} -\mathbf{AZ} \circ \widehat{\mathbf{W}}+ \mathbf{H} \right\|_F^2 +I_{+}(\mathbf{AZ}),
	\end{aligned}
\end{equation}
alternatingly for $\mathbf{X}$, $\mathbf{Z}$ and updating $\mathbf{H}$.
 $I_{+}(\mathbf{Q})$ takes $0$ if $\mathbf{Q} \succeq 0$ and $\infty$ otherwise. More specifically, let  $\mathbf{X}^{j}$, $\mathbf{Z}^{j}$ and $\mathbf{H}^{j}$ denote the variables at iteration $j$. Then, we have
\begin{equation}
	\begin{aligned}
		\mathbf{X}^{j+1} = \argmin_{\mathbf{X}} \frac{1}{2} \left\| \mathbf{RX} - \mathbf{S} \right\|_F ^2 + I_{+}(\mathbf{AX}) \\ + \frac{\lambda}{2} \left\| \mathbf{AX} -\mathbf{AZ}^{j} \circ \widehat{\mathbf{W}} + \mathbf{H}^{j} \right\|_F^2 ,\label{eq:forX}
	\end{aligned}
\end{equation}
\begin{equation}
	\mathbf{Z}^{j+1} = \argmin\limits_{\mathbf{Z}} \frac{\lambda}{2} \left\| \mathbf{AX}^{j+1} -\mathbf{AZ} \circ \widehat{\mathbf{W}} + \mathbf{H}^{j} \right\|_F^2  + I_{+}(\mathbf{AZ}),\label{eq:forZ}
\end{equation}
and update $\mathbf{H}^{j+1}$ by
\begin{equation}
	\mathbf{H}^{j+1} =\mathbf{H}^{j}+\mathbf{AX}^{j+1}-\mathbf{AZ}^{j+1} \circ \widehat{\mathbf{W}}.\label{eq:forP}
\end{equation}
A typical stopping criterion is to check whether $\mathbf{H}^{j}$ has stopped changing, i.e., ${\left\| \mathbf{H}^{j+1}-\mathbf{H}^{j} \right\|_F}/{\left\| \mathbf{H}^{j} \right\|_F} \leq \epsilon$ for a user-defined choice of $\epsilon$.
Problem~\eqref{eq:forX} can be written as
\begin{equation}
	\begin{aligned}
		\mathbf{X}^{j+1} = \argmin_{\mathbf{X}} \frac{1}{2} \left\| \left[ \begin{array}{c} \mathbf{R} \\ \lambda\mathbf{A} \end{array} \right] \mathbf{X} - \left[ \begin{array}{c} \mathbf{S} \\ \lambda\left(\mathbf{AZ}^{j} \circ \widehat{\mathbf{W}}  - \mathbf{H}^{j}\right)  \end{array} \right] \right\|_F^2 \\ ~\text{with}~ \mathbf{AX} \succeq 0,
	\end{aligned}
\end{equation}
and can be solved efficiently via quadratic programming.
Problem~\eqref{eq:forZ} can be solved efficiently using threholding, i.e., $\mathbf{Z}^{j+1}= \mathbf{A}^{-1} \left( \mathbf{AX}^{j+1} \circ \widehat{\mathbf{W}}^{-1} + \mathbf{H}^{j} \circ \widehat{\mathbf{W}}^{-1}  \right)$, if $\mathbf{AZ}^{j+1} \succeq 0$, and $\mathbf{Z}^{j+1} = \boldsymbol 0$ otherwise.

\subsection{AFODF Tractography}\label{sec:Tractography}
We introduce a tractography algorithm that caters to AFODFs using the direction getter mechanism described in \cite{garyfallidis2014dipy,amirbekian2016modeling}. The deterministic maximum direction getter returns
the most probable direction from a fiber orientation distribution.
When multiple plausible directions are available, the choice of best direction is dependent on the direction of the previous fiber segment.

\begin{figure*}[!t]\centering
	\includegraphics[width=0.8\textwidth]{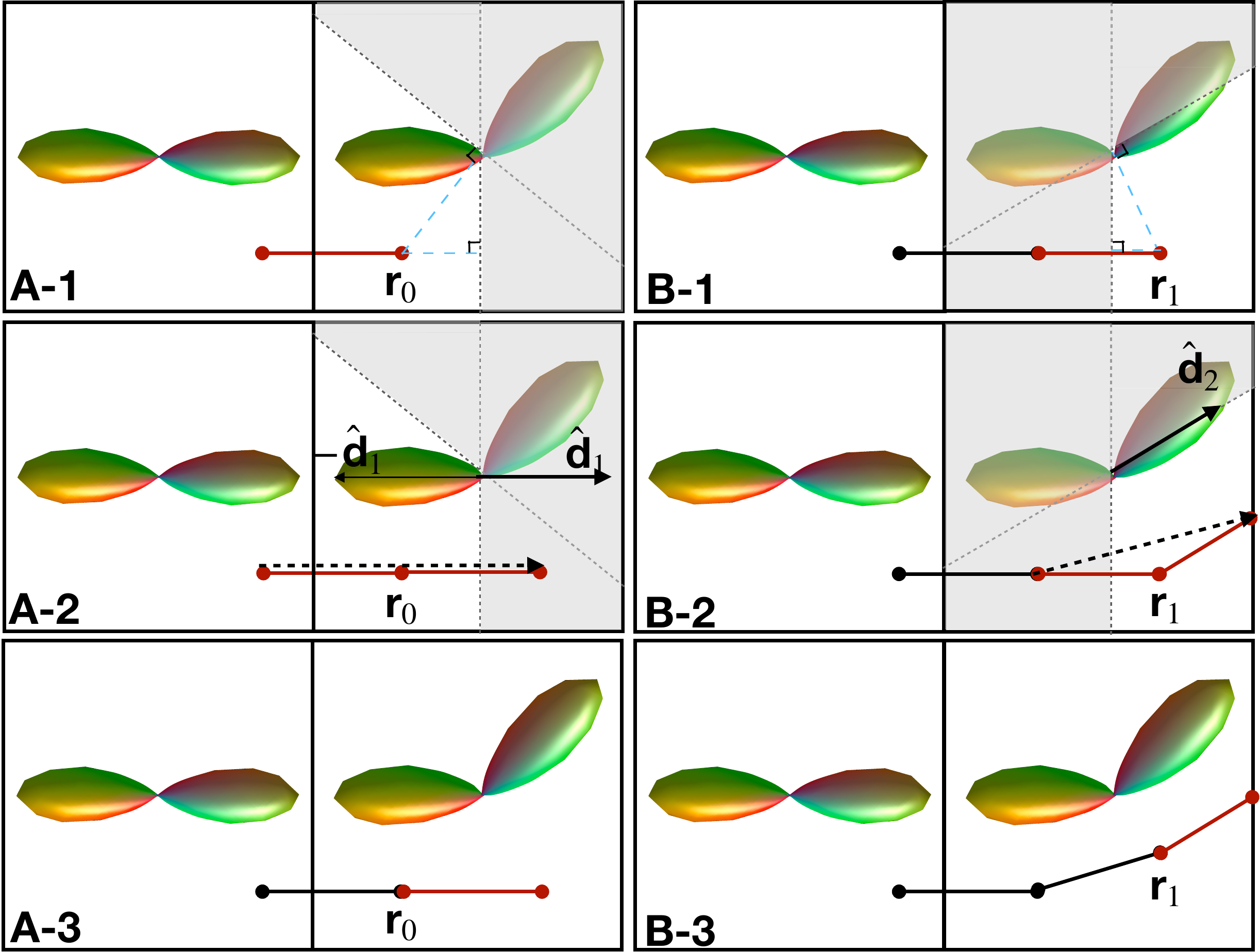}
	\caption{AFODFtractography algorithm for sharp sub-voxel bending. Tracking order: A-1, A-2, A-3, B-1, B-2, B-3.}\vspace{-10pt}
	\label{fig:tracking}
\end{figure*}

We extend the method described in \cite{bastiani2017improved} and the direction getter mechanism to work with AFODFs (Figure~\ref{fig:tracking}).
Similar to the deterministic fiber tracking method, we track along the most probable directions subject to the tracking constraints (e.g., maximum turning angle, anisotropy, etc.) \citep{amirbekian2016modeling}.
Unlike \cite{bastiani2017improved}, in our algorithm the direction may change within a voxel by following the most probable direction without referring to the direction in the previous voxel.  
The details of our multi-step algorithm are provided below (see Figure~\ref{fig:tracking}): 
\begin{itemize}
	\item Step A-1 (or B-1): Identify a sub-volume (non-shaded in Figure~\ref{fig:tracking}) based on the half-space defined by position ${\textbf{r}}_0$ (or ${\textbf{r}}_1$) and the plane through the voxel center that is perpendicular to ${\textbf{r}}_0$ (or ${\textbf{r}}_1$).

	\item Step A-2 (or B-2): Determine the most probable direction ($\hat{\textbf{d}}_1$ in A-2, $\hat{\textbf{d}}_2$ in B-2) in the sub-volume defined in Step A-1 (or B-1) for streamline propagation. The sign of direction is flipped whenever necessary.
   

	\item Step A-3 (or B-3): The direction of propagation and position are updated ($i=0$ for A-3, $i=1$ for B-3) using 
\begin{equation}
\hat{\textbf{d}}_i \leftarrow \frac{\hat{\textbf{d}}_i+\hat{\textbf{d}}_{i+1}}{\left\|\hat{\textbf{d}}_i+\hat{\textbf{d}}_{i+1}\right\|} ~~ \textrm{and}~~ \textbf{r}_i = \textbf{r}_{i-1} + s_{i}\hat{\textbf{d}}_i.
\end{equation}
%
Similar to \cite{bastiani2017improved}, the step size $s_{i}$ is determined by the multiplication of the AFODF value associated with $\textbf{d}_i$ and a scale $\rho$.
\end{itemize}


\subsection{Evaluation Metrics}\label{sec:Evaluation}

\subsubsection{Asymmetry Index (ASI)}
Based on \cite{karayumak2018asymmetric}, we define an asymmetry index that measures the difference between $F(\mathbf{u})$ and $F(-\mathbf{u})$ for a set of directions $\mathbb{U} = \{\mathbf{u}_{1},\mathbf{u}_{2},\hdots,\mathbf{u}_{K}\}$:
\begin{equation}
\text{ASI} = \sqrt{1-\left(  \frac{\sum_{\mathbf{u} \in \mathbb{U} } F(\mathbf{u})F(-\mathbf{u})}{\sum_{\mathbf{u} \in \mathbb{U} } F(\mathbf{u})^2} \right)^2},
\end{equation}
which has a value of 0 when $F(\mathbf{u})=F(-\mathbf{u})$ for all $\mathbf{u}\in\mathbb{U}$.

\subsubsection{Model Discrepancy Index (MDI)}
A model discrepancy indices (MDI) is used to measure the distance between an AFODF and its symmetric counterpart.
A Chebyshev distance is used to measure the greatest FODF difference on a unit sphere $\mathbb{S}^{2}$.
Let $\hat{F}_1$ and $\hat{F}_2$ be the two kinds of FODFs, e.g., asymmetric and symmetric, which are normalized to range $[0, 1]$. Then, the MDI is defined as
\begin{equation}
\text{MDI} = \lim\limits_{p \rightarrow \infty} \left(   \sum_{\mathbf{u} \in \mathbb{U} } \left| \hat{F}_{1}(\mathbf{u})  - \hat{F}_{2}(\mathbf{u}) \right|^p   \right)^{1/p},\label{eq:che}
\end{equation}
which ranges from 0 to 1.

\subsubsection{Interscan Consistency Index (ICI)}
Based on \cite{streiner2003starting}, we introduce an interscan consistency index (ICI) to measure the similarity of the distributions of the streamline endpoints
of the same subject across two different scans (e.g., acquired using 3T and 7T scanners). Given two vectors, $\mathbf{c}_{1}=[c_{1,1},\hdots,c_{1,n}]$ and $\mathbf{c}_{2}=[c_{2,1},\hdots,c_{2,n}]$, that represent the fractions of endpoints in $n$ regions of interest (ROIs), the index is defined as
\begin{equation}
\text{ICI} = \left| \frac{\sigma_{{\mathbf{c}}_{1},{\mathbf{c}}_{2}}}{\sigma_{{\mathbf{c}}_{1}}\sigma_{{\mathbf{c}}_{2}}} \right|, \label{eq:ICI}
\end{equation}
where
\begin{equation*}
\begin{aligned}
\sigma_{ {\mathbf{c}}_{1}, {\mathbf{c}}_{2} }
&= \frac{1}{n}\sum_{i=1}^{n} \left( {c}_{1,i} - \bar{c}_{1} \right) \left( {c}_{2,i} - \bar{c}_{2} \right),\\~
\bar{c}_{k} &= \frac{1}{n}\sum_{i=1}^{n} {c}_{k,i},~\sigma_{ {\mathbf{c}}_{k} } := \sigma_{ {\mathbf{c}}_{k}, {\mathbf{c}}_{k} }.
\end{aligned}
\end{equation*}
ICI ranges from 0 (inconsistent) to 1 (consistent). Note that ICI is quite general and can be used to measure the interscan consistency of quantities other than endpoint statistics. 

\section{Experimental Results}\label{sec:Experiments}

\subsection{Materials}
To demonstrate the advantages of AFODFs, we utilized the 3T and 7T DMRI dataset (see Table~\ref{tab:HCP}) of a subject (\# 105923) scanned in the Human Connectome Project (HCP) \citep{van2013wu}.
More acquisition details are given in \cite{sotiropoulos2016fusion,van2013wu}.

\begin{table}[h]\scriptsize\centering
	\caption{Summary of the HCP 3T and 7T dMRI protocols.}
	\begin{tabular}{lll}
		\hline
		&  \textbf{HCP 3T} &  \textbf{HCP 7T}\tabularnewline
		\hline
		Spatial resolution & (1.25\,mm)$^3$, LR/RL phase encoding & (1.05\,mm)$^3$, AP/PA phase encoding\tabularnewline
		Acceleration &  Multiband = 3 &  Multiband = 2 , GRAPPA = 3\tabularnewline
		Total echo train length &  84.24\,ms &  41\,ms\tabularnewline
		Gradient strength (max) & 100\,mT/m & 70\,mT/m\tabularnewline
		b-value (s/mm$^2$) & 1000, 2000, 3000 & 1000, 2000\tabularnewline
		Gradient directions & 270 & 130 \tabularnewline
		\hline
	\end{tabular}
	\label{tab:HCP}
\end{table}

Tissue segmentation (CSF, cortical GM, deep GM, and WM) was performed using the T1-weighted image using the pipeline described in \cite{smith2012anatomically}. 
The pipeline was performed using MRtrix software package \citep{tournier2012mrtrix}.
A mask generated based on the cortical and deep GM volume fraction maps was warped to the space of the diffusion-weighted images for tractography seeding. The cortical regions were labeled based on the Destrieux atlas \citep{destrieux2010automatic}.

According to \cite{Jeurissen2014}, a fiber RF is estimated from the diffusion signal for each $b$-shell and each tissue type.
The WM RF is estimated by averaging the reoriented diffusion signal profiles of voxels with high anisotropy.
The isotropic RFs for GM and CSF are estimated from isotropic representative voxels of GM and CSF, respectively.

We also utilized a macaque diffusion MRI dataset with histological data\footnote{https://www.nitrc.org/projects/e39macaque} for validation. The brain was placed in liquid Fomblin (California Vacuum Technology, CA) and scanned on a Varizan 9.4\,T, 21\,cm bore magnet. The structural image was acquired using a 3D gradient echo sequence (TR=50\,ms; TE=3\,ms; flip angle=45$^\circ$) at 200 $\mu$m isotropic resolution. The diffusion data were acquired with a 3D spin-echo diffusion-weighted EPI sequence at 400\,$\mu$m isotropic resolution. The b-values were set to 3000, 6000, 9000, and 12000\,s/mm$^2$. These values were chosen due to the decreased diffusivity of ex-vivo tissue, which is approximately triple of that in vivo, and is expected to closely replicate the signal attenuation profile for in vivo tissue with a b-values of 1000, 2000, 3000, and 4000\,s/mm$^2$. A gradient table of 404 uniformly distributed directions was used to acquire diffusion-weighted volumes with 16 additional non-diffusion-weighted volumes collected at b=0\,s/mm$^2$. More acquisition details can be found in \cite{schilling2018confirmation}.

\begin{table}[t]\scriptsize\centering
	\caption{Abbreviations of gyral blades \citep{schilling2018confirmation}.}
		\begin{tabular}{cccccc}
			\hline
			Abbreviation & Name & Abbreviation & Name \tabularnewline
			\hline
			SFG & superior frontal gyrus & MFG & medial frontal gyrus \tabularnewline
			IFG & inferior frontal gyrus &	FOG & frontal orbital gyrus \tabularnewline
			LorG & lateral orbital gyrus & MorG & medial orbital gyrus\tabularnewline
			Gre & gyrus rectus & ACgG & anterior cingulate gyrus \tabularnewline
			PrG & pre- central gyrus & STG & superior temporal gyrus  \tabularnewline
			INS & insula & MTG & middle temporal gyrus \tabularnewline
			ITG & inferior temporal gyrus & PoG & postcentral gyrus \tabularnewline
			PCgG & posterior cingulate gyrus & SMG & supramarginal gyrus \tabularnewline
			FuG & fusiform gyrus & PPhG & posterior parahippocampal gyrus \tabularnewline
			SPL & superior parietal lobule & AnG & angular gyrus \tabularnewline
			IOG & inferior occipital gyrus & LiG & lingual gyrus \tabularnewline
			CUN & cuneus & OG & occipital gyrus\tabularnewline
			\hline
		\end{tabular}
	\label{tab:cortical_label}
\end{table}

\subsection{Macaque Results}
\begin{figure*}[!t]\centering
	\begin{tabular}{m{0.4\textwidth}m{0.3\textwidth}m{0.3\textwidth}}
    & \centerline{\textbf{FODF}} & \centerline{\textbf{AFODF}}\\[-20pt]
		\begin{minipage}{0.6\textwidth}
			\includegraphics[width=0.7\textwidth]{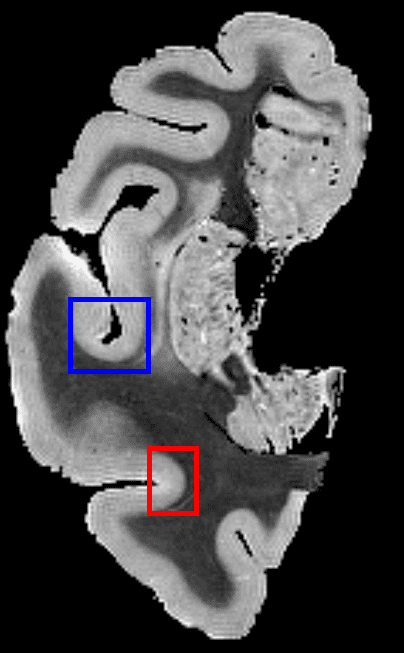}
		\end{minipage}&
		\begin{minipage}{0.5\textwidth}
			\includegraphics[width=0.6\textwidth]{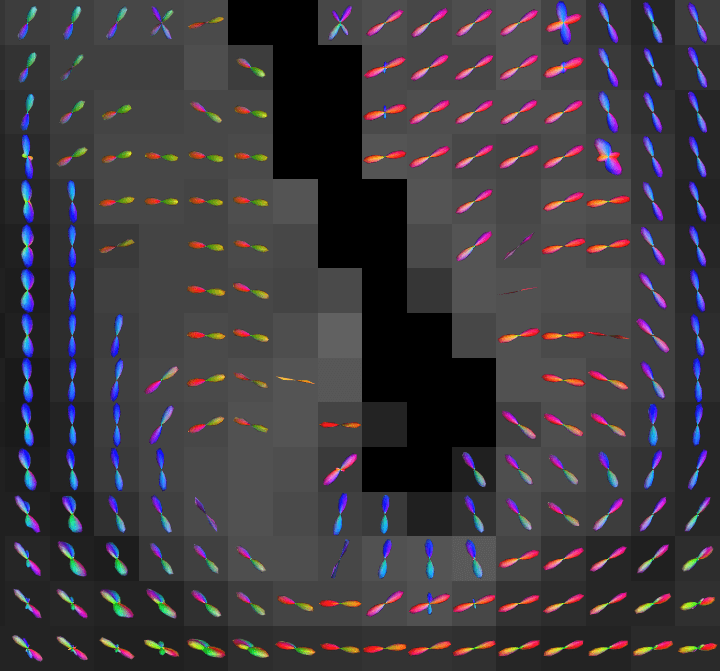}\\[1ex]
			\includegraphics[width=0.6\textwidth]{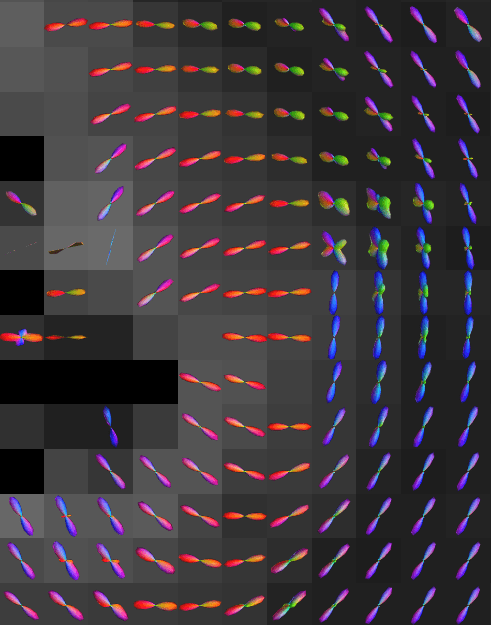}
		\end{minipage}&
		\begin{minipage}{0.5\textwidth}
			\includegraphics[width=0.6\textwidth]{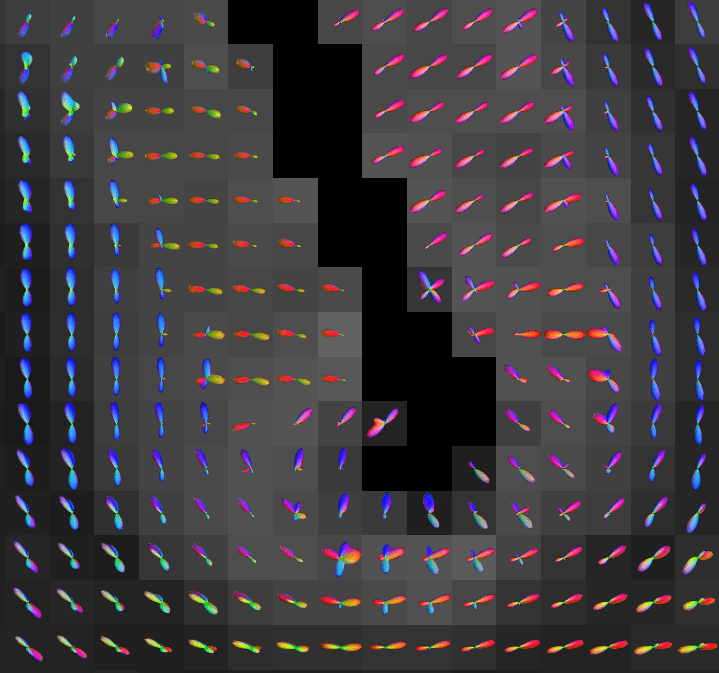}\\[1ex]
			\includegraphics[width=0.6\textwidth]{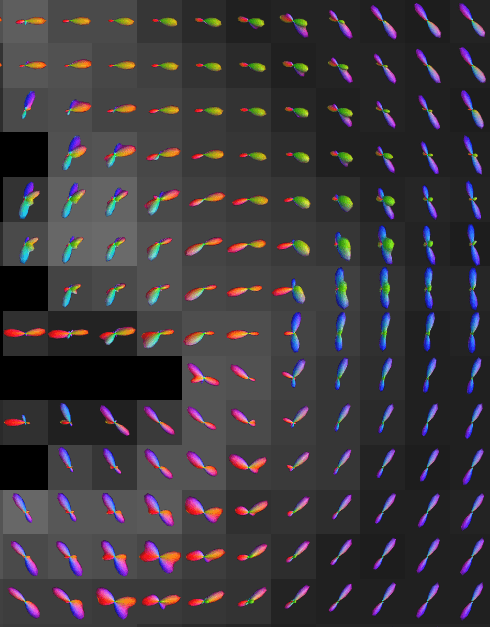}
		\end{minipage}
	\end{tabular}
	
	\caption{AFDOFs capture the bending and branching fiber populations at the WM-GM boundary. Orientations are mostly perpendicular to the WM-GM surface.}\vspace{-10pt}
	
	\label{fig:monkey_odf}
	
\end{figure*}

\begin{figure*}[!t]\centering
	\begin{tabular}{m{0.415\textwidth}m{0.01\textwidth}m{0.415\textwidth}}
    & & 
    \\[-20pt]
		\begin{minipage}{0.6\textwidth}
			\includegraphics[width=0.7\textwidth]{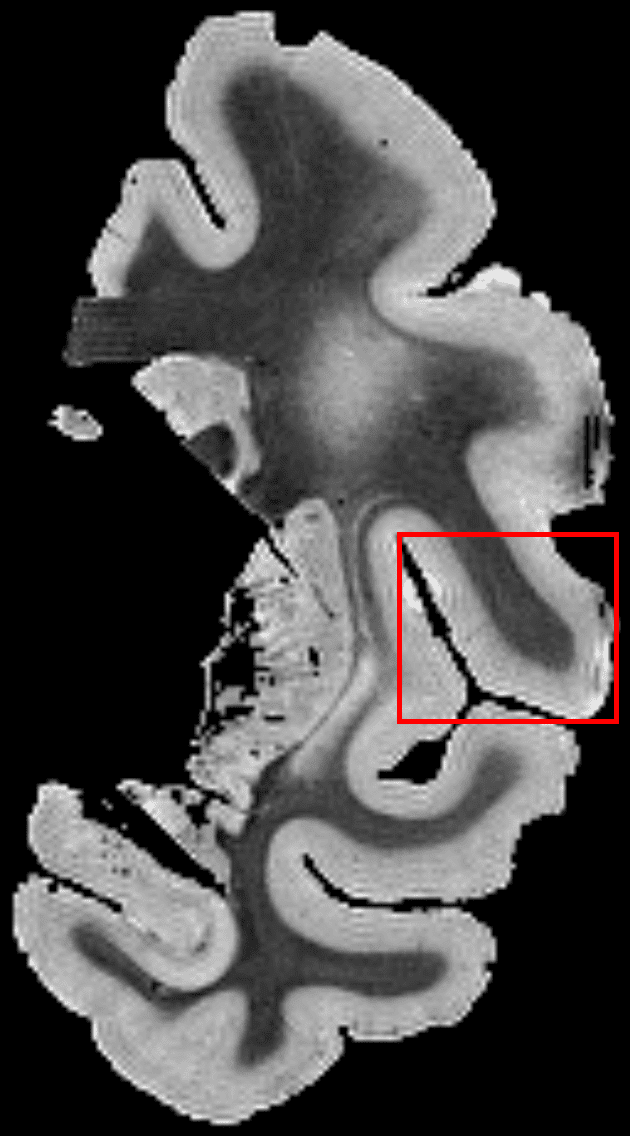}
		\end{minipage}&
		\begin{minipage}{0.01\textwidth}
			\rotatebox{90}{\hspace{40pt}\centerline{\textbf{FODF}}}\\[1ex]
			\rotatebox{90}{\hspace{-60pt}\centerline{\textbf{AFODF}}}
		\end{minipage}&
		\begin{minipage}{0.415\textwidth}
			\includegraphics[width=\textwidth]{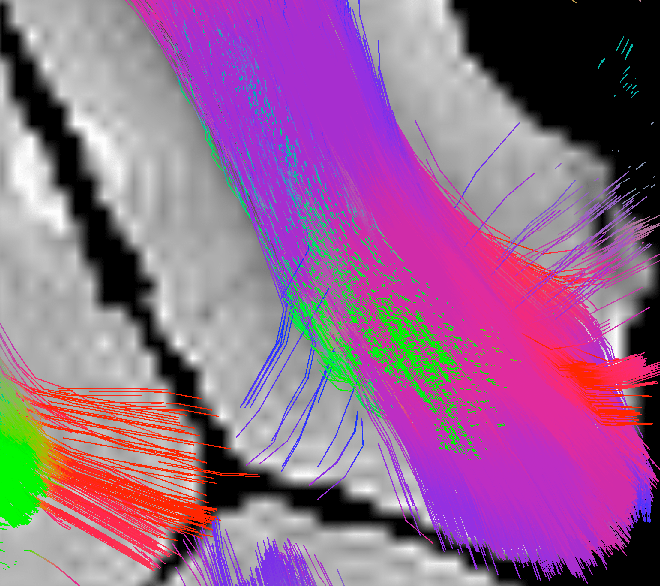}\\[1ex]
			\includegraphics[width=\textwidth]{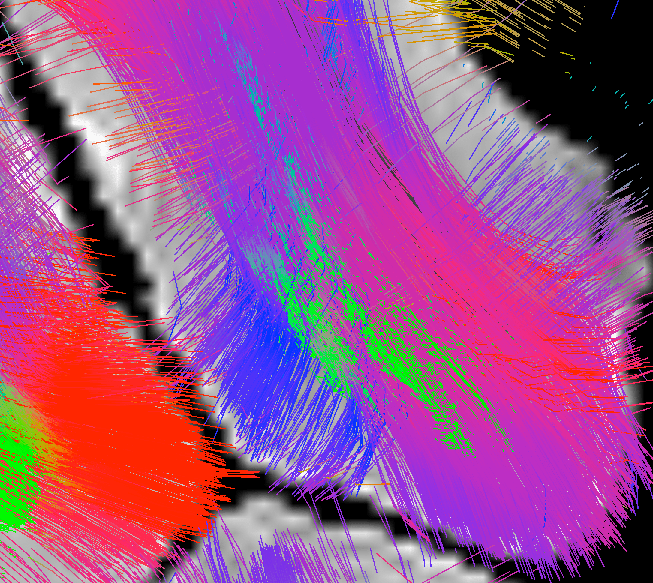}
		\end{minipage}
	\end{tabular}
	
	\caption{Fiber streamlines at a gyral blade, generated with WM seeds.}
\label{fig:monkey_fiber}
\end{figure*}

Figure~\ref{fig:monkey_odf} shows the intravoxel architecture as given by FODFs and AFODFs in sulcal banks and fundi. 
Many areas in both WM and GM show multiple fiber populations. 
In agreement with previous studies \citep{leuze2012layer}, we see that AFODFs are largely oriented perpendicularly to the WM-GM boundary at the sulcal banks.
Tractography results in Figure~\ref{fig:monkey_fiber} confirm this observation, validating that AFODFs improve tractography across WM-GM boundaries. 
This increases cortical coverage and reduces gyral bias.

\begin{figure*}[!th]\centering
	\begin{tabular}{m{0.33\textwidth}m{0.33\textwidth}m{0.33\textwidth}m{0.33\textwidth}}
     & \centerline{\textbf{Gyral Blades}} & \centerline{\textbf{Gyri}} \\ [-20pt]
		\begin{minipage}{0.33\textwidth}
			\includegraphics[width=\textwidth]{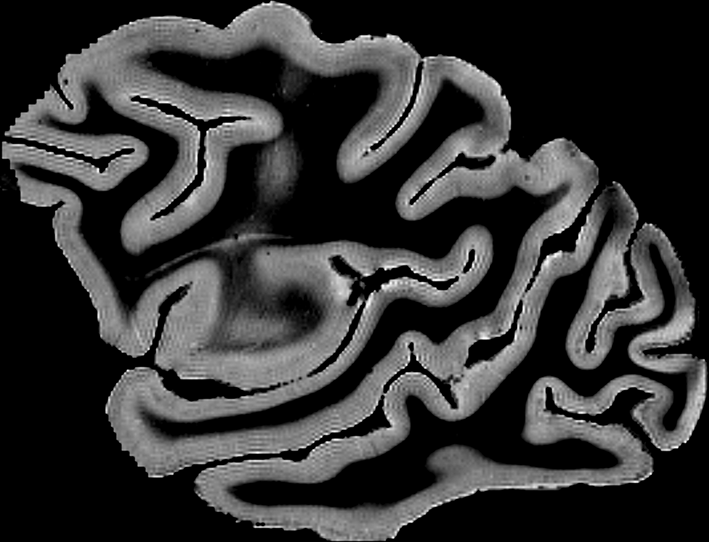}
		\end{minipage}&
		\begin{minipage}{0.33\textwidth}
			\includegraphics[width=\textwidth]{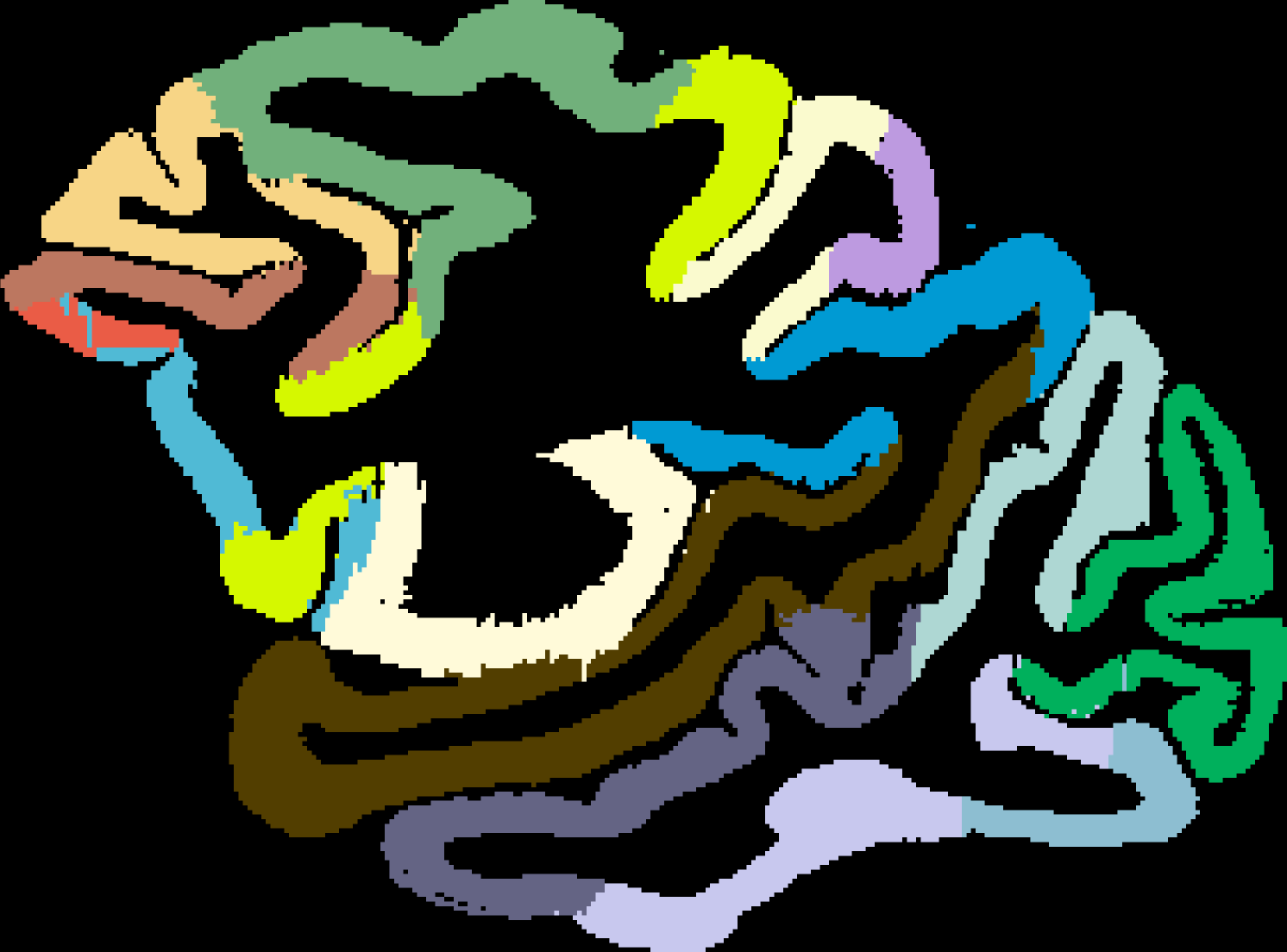}
		\end{minipage}&
		\begin{minipage}{0.33\textwidth}
			\includegraphics[width=\textwidth]{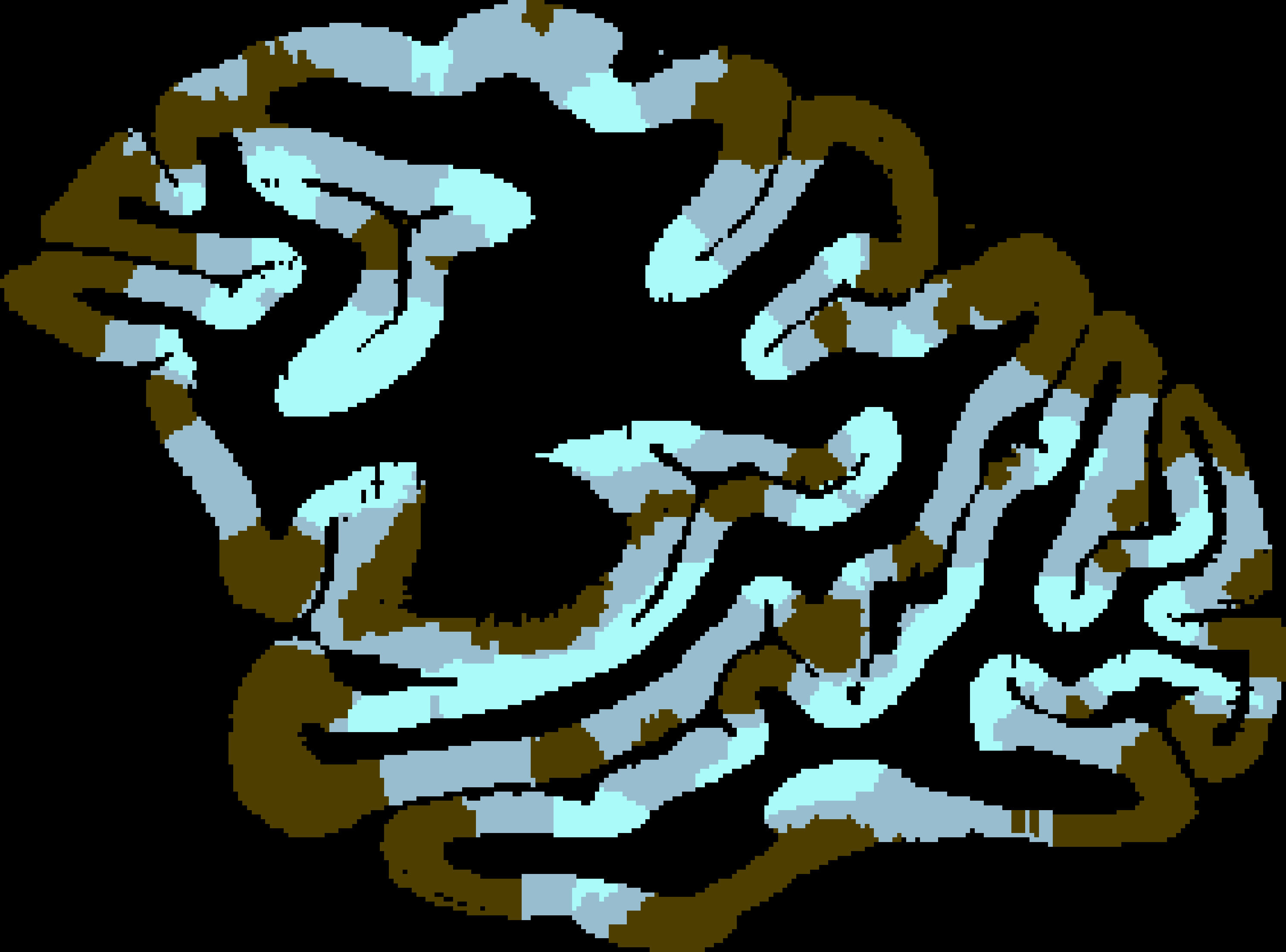}
		\end{minipage}\\
	\end{tabular}
	\caption{Gyral blades and gyri labeled based on the structural image. The brown areas and the adjacent lake blue areas merge into gyri.}
	\label{fig:structure}
\end{figure*}

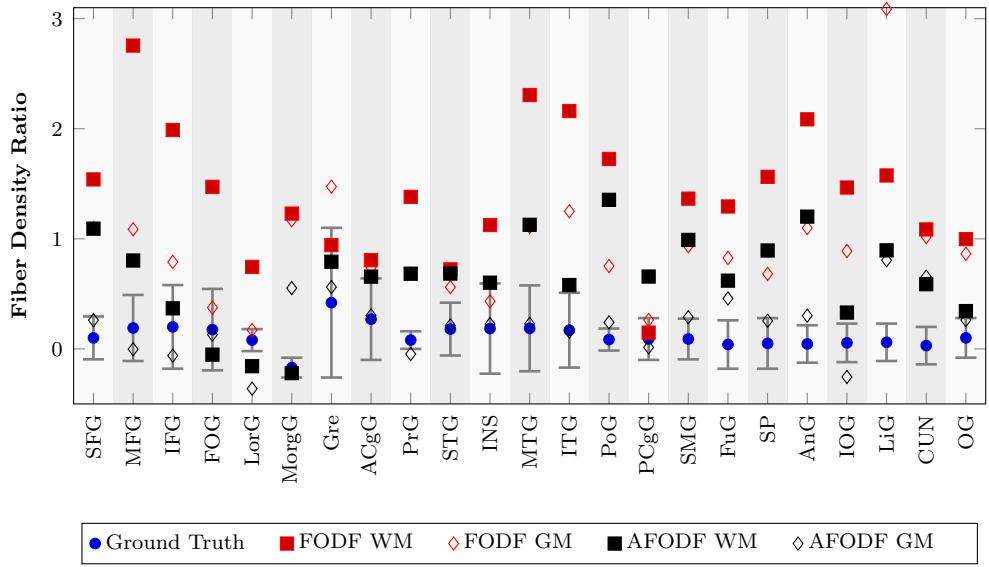
\begin{figure}[!ht]
	\centering
	\pgfplotsset{width=\textwidth,height=0.5\textwidth,compat=1.13}
	
	\begin{tikzpicture}
	\begin{axis}[
	ymin=-0.5,ymax=3.1,
	xmin=0.5,xmax=23.5,
	ylabel={\textbf{Fiber Density Ratio}},
	xtick=data,
	xticklabels={SFG,MFG,IFG,FOG,LorG,MorgG,Gre,ACgG,PrG,STG,INS,MTG,ITG,PoG,PCgG,SMG,FuG,SP,AnG,IOG,LiG,CUN,OG},
	x tick label style={rotate=90,anchor=east},
	axis on top,
	legend style={at={(0.5, -0.30)}, anchor=north, legend columns=5, cells={anchor=center, fill}}
	]

	\addplot+[only marks,
	error bars/.cd,
	y dir=both,
	y explicit,
	error bar style={line width=1pt,gray},
	error mark options={
		rotate=90,
		gray,
		mark size=4pt,
		line width=1pt
	}
	]
	table[x=X,y=GT_mean, y error=GT_std]{plot/monkey_FD.txt};\addlegendentry{Ground Truth~~~~~}
	
	\addplot+[only marks,mark size=1pt,mark=square*,mark size=2.5pt, color=red] table[x=X,y=FODF_DET_seedWM]{plot/monkey_FD.txt};\addlegendentry{FODF WM~~~~~}
	\addplot+[only marks,,mark=diamond,mark size=2.5pt,color=red] table[x=X,y=FODF_DET_seedGM]{plot/monkey_FD.txt};\addlegendentry{FODF GM~~~~~}
	\addplot+[only marks,mark size=1pt,mark=square*,mark size=2.5pt,color=black] table[x=X,y=AFODF_DET_seedWM]{plot/monkey_FD.txt};\addlegendentry{AFODF WM~~~~~}
	\addplot+[only marks,mark size=1pt,mark=diamond,mark size=2.5pt,color=black] table[x=X,y=AFODF_DET_seedGM]{plot/monkey_FD.txt};\addlegendentry{AFODF GM~~~~~}
	
	\fill [gray, fill opacity=0.05] (axis cs:0.5,-0.5) rectangle (axis cs:1.5,3.1);
	\fill [gray, fill opacity=0.15] (axis cs:1.5,-0.5) rectangle (axis cs:2.5,3.1);
	\fill [gray, fill opacity=0.05] (axis cs:2.5,-0.5) rectangle (axis cs:3.5,3.1);
	\fill [gray, fill opacity=0.15] (axis cs:3.5,-0.5) rectangle (axis cs:4.5,3.1);
	\fill [gray, fill opacity=0.05] (axis cs:4.5,-0.5) rectangle (axis cs:5.5,3.1);
	\fill [gray, fill opacity=0.15] (axis cs:5.5,-0.5) rectangle (axis cs:6.5,3.1);
	\fill [gray, fill opacity=0.05] (axis cs:6.5,-0.5) rectangle (axis cs:7.5,3.1);
	\fill [gray, fill opacity=0.15] (axis cs:7.5,-0.5) rectangle (axis cs:8.5,3.1);
	\fill [gray, fill opacity=0.05] (axis cs:8.5,-0.5) rectangle (axis cs:9.5,3.1);
	\fill [gray, fill opacity=0.15] (axis cs:9.5,-0.5) rectangle (axis cs:10.5,3.1);
	\fill [gray, fill opacity=0.05] (axis cs:10.5,-0.5) rectangle (axis cs:11.5,3.1);
	\fill [gray, fill opacity=0.15] (axis cs:11.5,-0.5) rectangle (axis cs:12.5,3.1);
	\fill [gray, fill opacity=0.05] (axis cs:12.5,-0.5) rectangle (axis cs:13.5,3.1);
	\fill [gray, fill opacity=0.15] (axis cs:13.5,-0.5) rectangle (axis cs:14.5,3.1);
	\fill [gray, fill opacity=0.05] (axis cs:14.5,-0.5) rectangle (axis cs:15.5,3.1);
	\fill [gray, fill opacity=0.15] (axis cs:15.5,-0.5) rectangle (axis cs:16.5,3.1);
	\fill [gray, fill opacity=0.05] (axis cs:16.5,-0.5) rectangle (axis cs:17.5,3.1);
	\fill [gray, fill opacity=0.15] (axis cs:17.5,-0.5) rectangle (axis cs:18.5,3.1);
	\fill [gray, fill opacity=0.05] (axis cs:18.5,-0.5) rectangle (axis cs:19.5,3.1);
	\fill [gray, fill opacity=0.15] (axis cs:19.5,-0.5) rectangle (axis cs:20.5,3.1);
	\fill [gray, fill opacity=0.05] (axis cs:20.5,-0.5) rectangle (axis cs:21.5,3.1);
	\fill [gray, fill opacity=0.15] (axis cs:21.5,-0.5) rectangle (axis cs:22.5,3.1);
	\fill [gray, fill opacity=0.05] (axis cs:22.5,-0.5) rectangle (axis cs:23.5,3.1);

	\end{axis}
	\end{tikzpicture}
	\caption{
    The histological ground truth density ratios are shown as blue circles (mean $\pm$ 95\% confidence interval). Results for Both WM seeding GM seeding are shown. Logarithmic scale is used for the vertical axis.}
	\label{fig:monkey_FD}
\end{figure}

Figure~\ref{fig:monkey_FD} shows the ratios of average fiber densities of gyral crowns to sulcal banks for the gyral blades listed in Table~\ref{tab:cortical_label} and shown in Figure~\ref{fig:structure}, using WM and GM seeding. 
Fiber density is computed as the number of streamlines per voxel.
FODF tractography is biased towards gyral crowns in comparison with the axonal density ratios given by the histological data in many gyral blades, for both seeding strategies. On the other hand, AFODF tractography yields density ratios that match the histological data better.

\par

\subsection{Intravoxel Architecture}
\begin{figure*}[t]\centering
	\begin{tabular}{m{0.01\textwidth}m{0.4\textwidth}|m{0.4\textwidth}}
		& \centerline{\textbf{FODF}} & \centerline{\textbf{AFODF}}\\[-5pt]
		
		\rotatebox{90}{\centerline{\textbf{3T}}}
		& \includegraphics[width=0.4\textwidth]{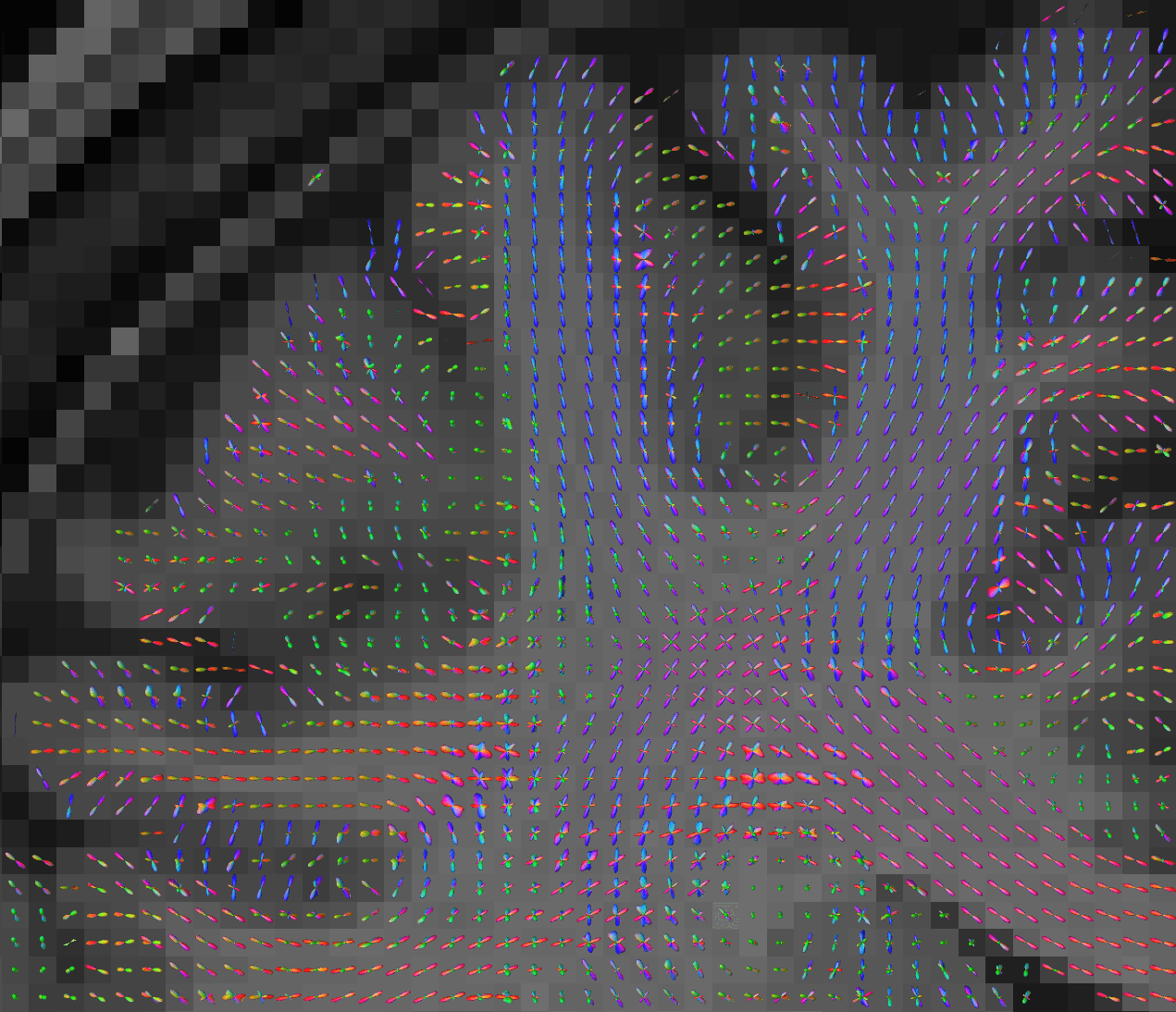} & \scalebox{-1}[1]{\includegraphics[width=0.4\textwidth]{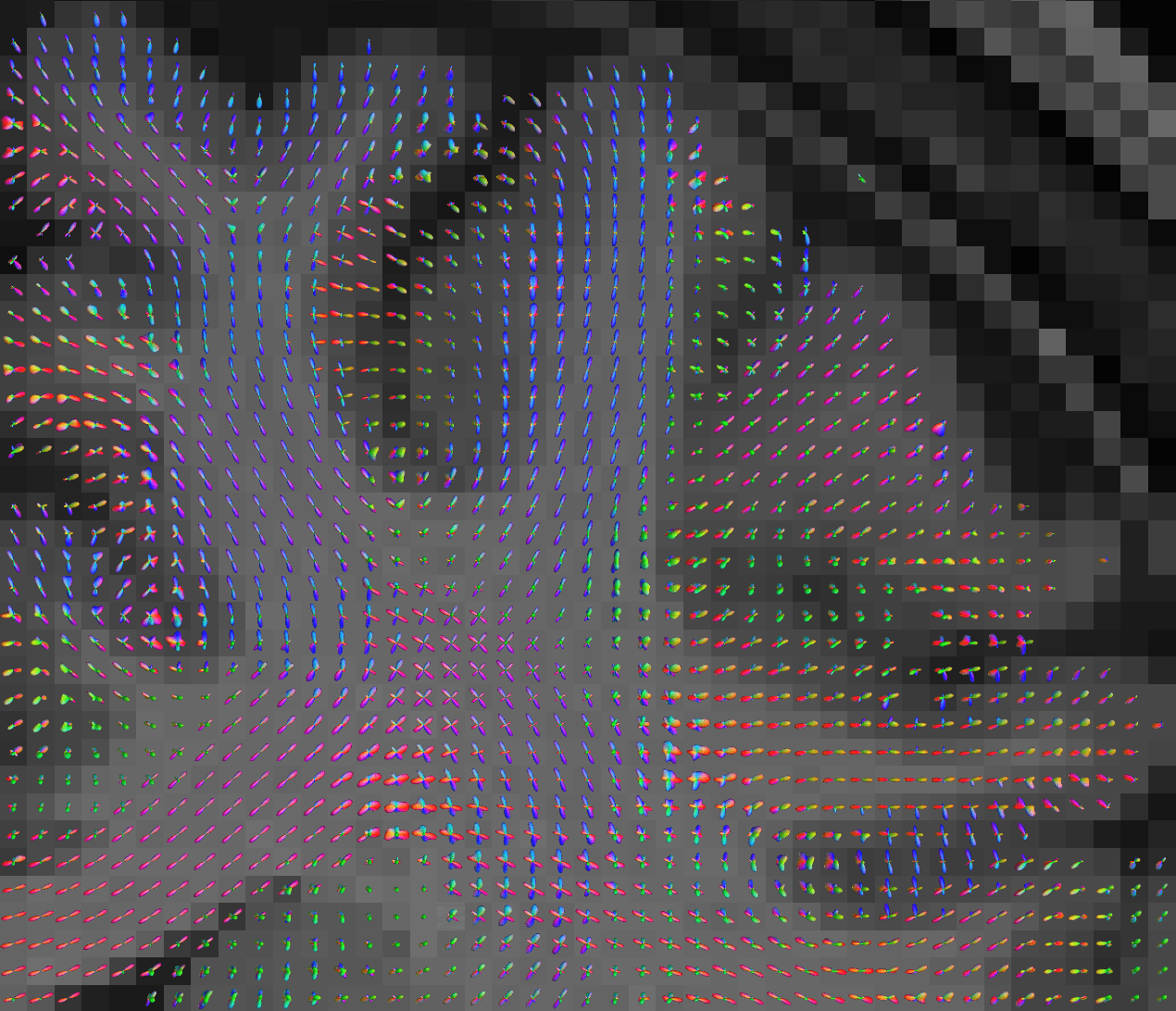}}\\
		
		\rotatebox{90}{\centerline{\textbf{7T}}}
		& \includegraphics[width=0.4\textwidth]{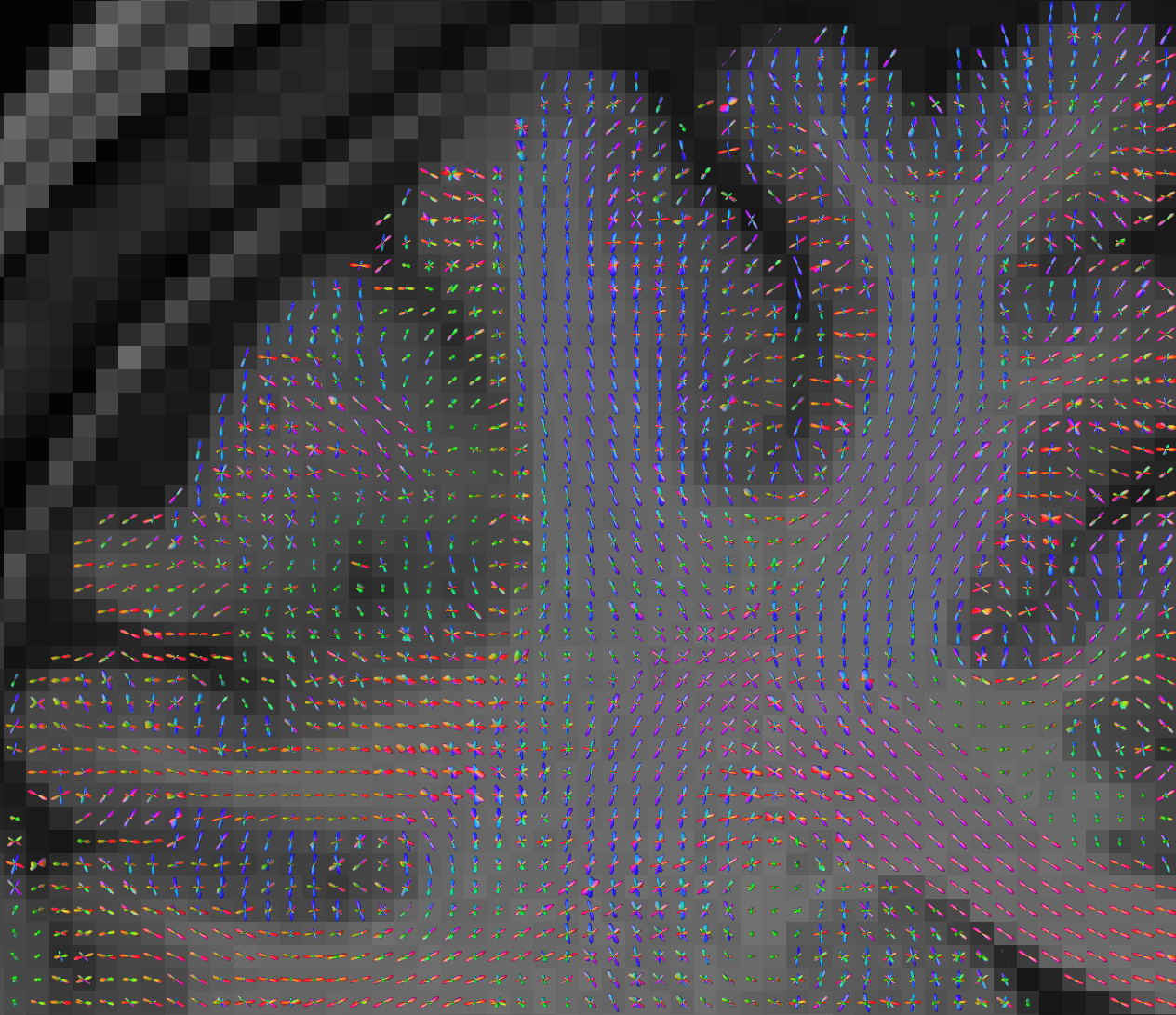} & \scalebox{-1}[1]{\includegraphics[width=0.4\textwidth]{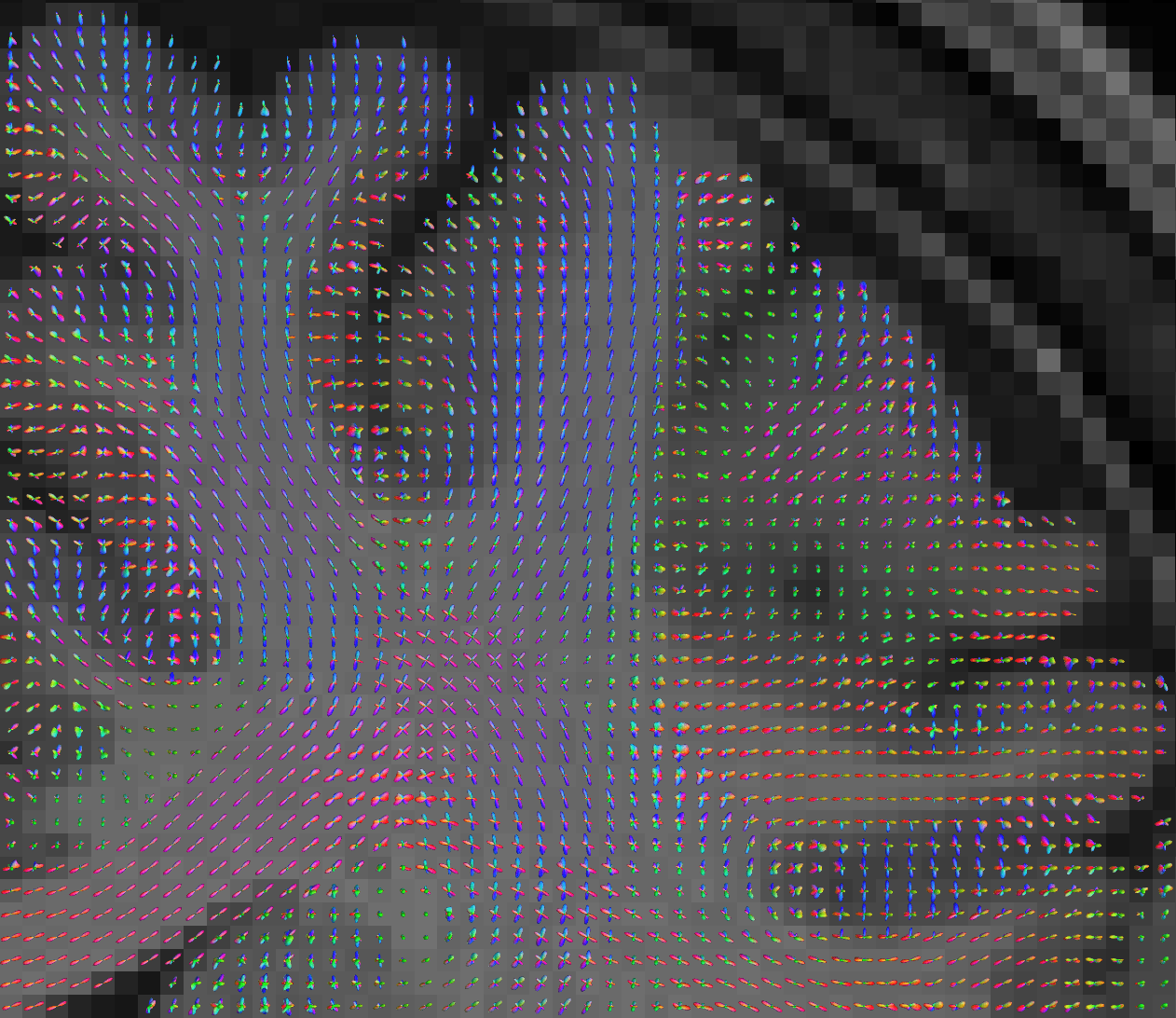}}\\
	\end{tabular}
	
	\caption{Min-max normalized FODFs and AFODFs.}
	\label{fig:ODF_largeROI}
\end{figure*}

\begin{figure*}[!t]
	
	\begin{tabular}{m{0.001\textwidth}m{0.2\textwidth}m{0.2\textwidth}m{0.22\textwidth}m{0.22\textwidth}}
		& \centerline{\textbf{FODF}} & \centerline{\textbf{AFODF}}& \centerline{\textbf{FODF}} & \centerline{\textbf{AFODF}}\\[-10pt]
		
		\rotatebox{90}{\centerline{\hspace{15pt}\textbf{3T}}}
		&
		\includegraphics[height=0.23\textwidth]{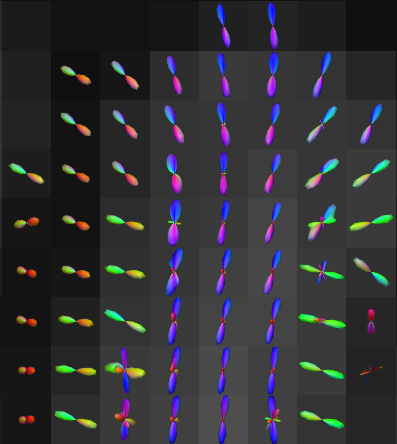}
		&
		\includegraphics[height=0.23\textwidth]{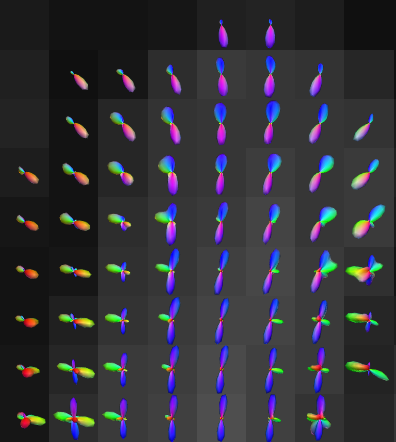}
		&
		\includegraphics[height=0.23\textwidth]{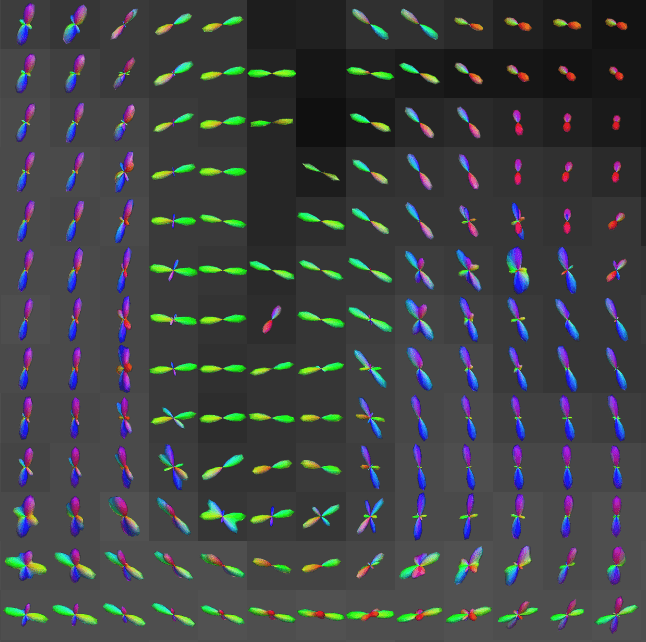}
		&
		\includegraphics[height=0.23\textwidth]{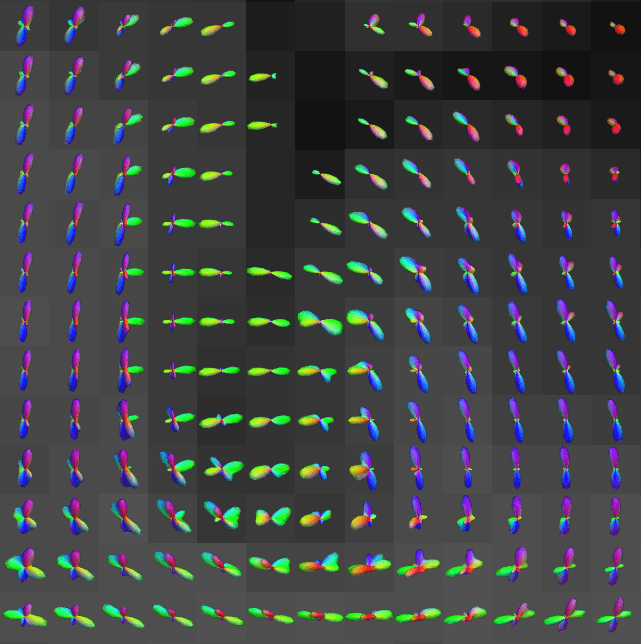}
		\\
		
		\rotatebox{90}{\centerline{\hspace{15pt}\textbf{7T}}}
		&
		\includegraphics[height=0.23\textwidth]{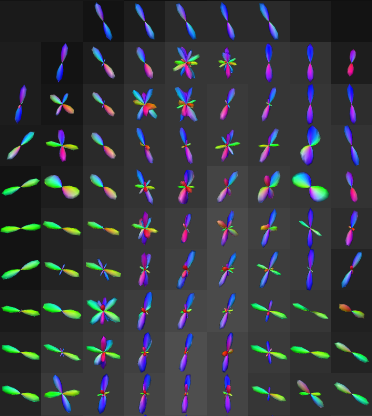}
		&
		\includegraphics[height=0.23\textwidth]{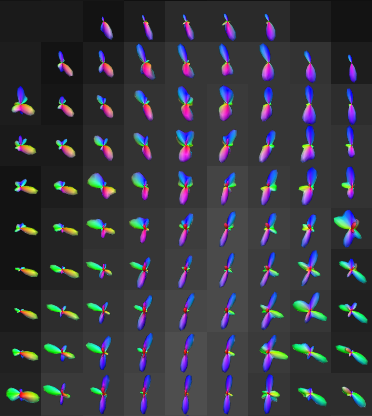}
		&
		\includegraphics[height=0.23\textwidth]{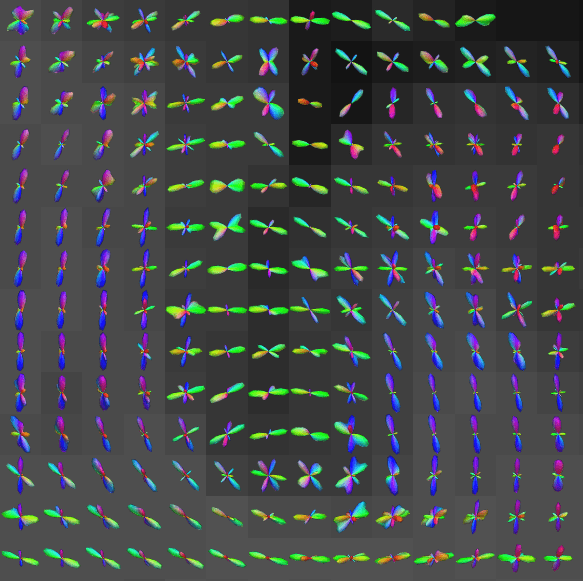}
		&
		\includegraphics[height=0.23\textwidth]{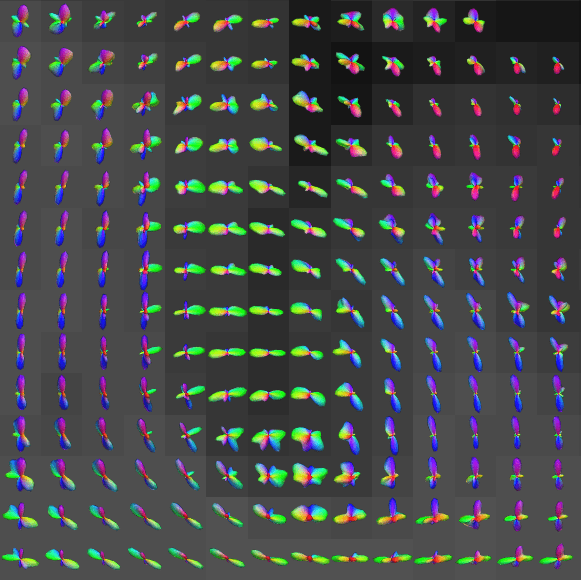}
		\\
	\end{tabular}
	
	\caption{Close-up views of FODFs and AFODFs in gyral blades.}
	\label{fig:ODF_SmallROI}
\end{figure*}

\begin{figure*}[!t]\centering
	
	\setlength{\colorbarlength}{0.12\textwidth}
	
	\pgfdeclarehorizontalshading{spectrum1}{0.12\colorbarlength}
	{
		rgb(0)=(0,0,0); rgb(0.10\colorbarlength)=(0,0,1); rgb(0.33\colorbarlength)=(0,1,1); rgb(0.5\colorbarlength)=(0,1,0); rgb(0.66\colorbarlength)=(1,1,0);
		rgb(1.0\colorbarlength)=(1,0,0)
	}
	
	\pgfdeclarehorizontalshading{spectrum2}{0.12\colorbarlength}
	{
		rgb(0)=(0,0,0); rgb(1.0\colorbarlength)=(1,1,1)
	}
	
	\newcommand\colorBar[3]{		
		\begin{tikzpicture}
		\node[inner sep=0] (#3)
		{
			\pgfuseshading{#3}
		};
		\node [left=0mm of #3] {\scriptsize{#1}};
		\node [right=-0.5mm of #3] {\scriptsize{#2}};
		\end{tikzpicture}
	}
	
	\begin{tabular}{m{0.01\textwidth}m{0.24\textwidth}m{0.20\textwidth}m{0.24\textwidth}m{0.20\textwidth}}
		& \multicolumn{2}{c}{\textbf{3T}} & \multicolumn{2}{c}{\textbf{7T}}\\
		\rotatebox{90}{\centerline{\textbf{ASI}}} &
		\begin{minipage}{0.3\textwidth}
			\includegraphics[width=0.7\textwidth]{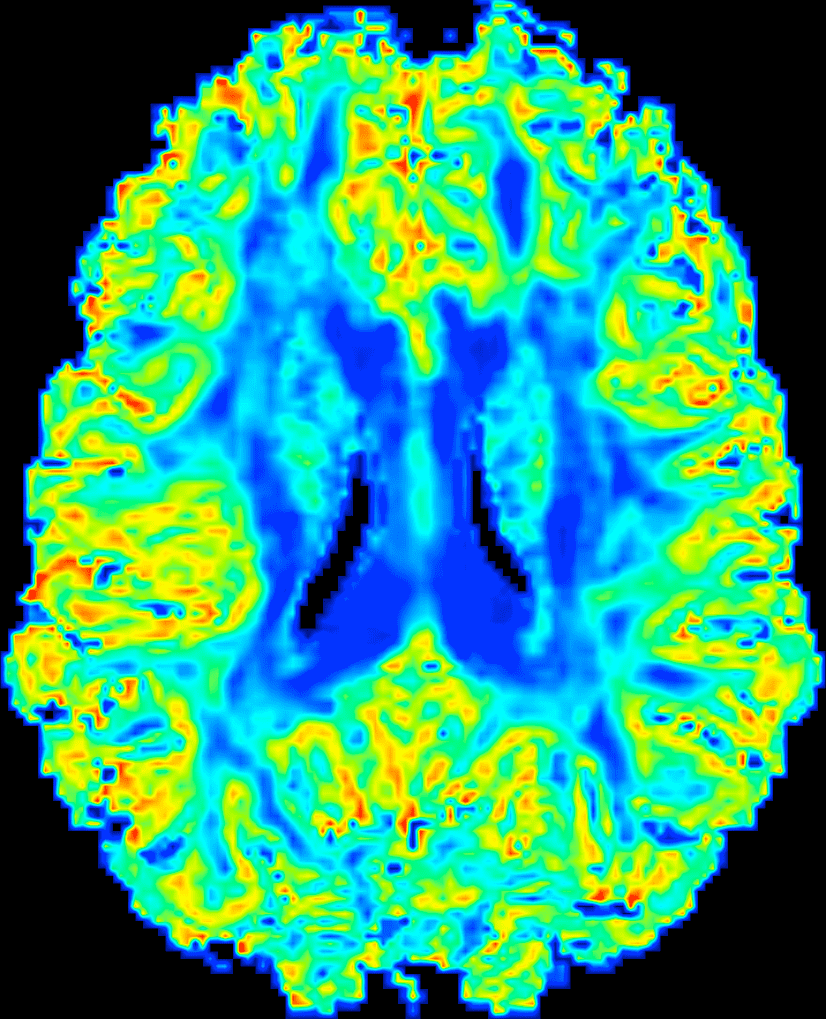}
		\end{minipage}&
		\begin{minipage}{0.2\textwidth}
			\includegraphics[width=0.8\textwidth]{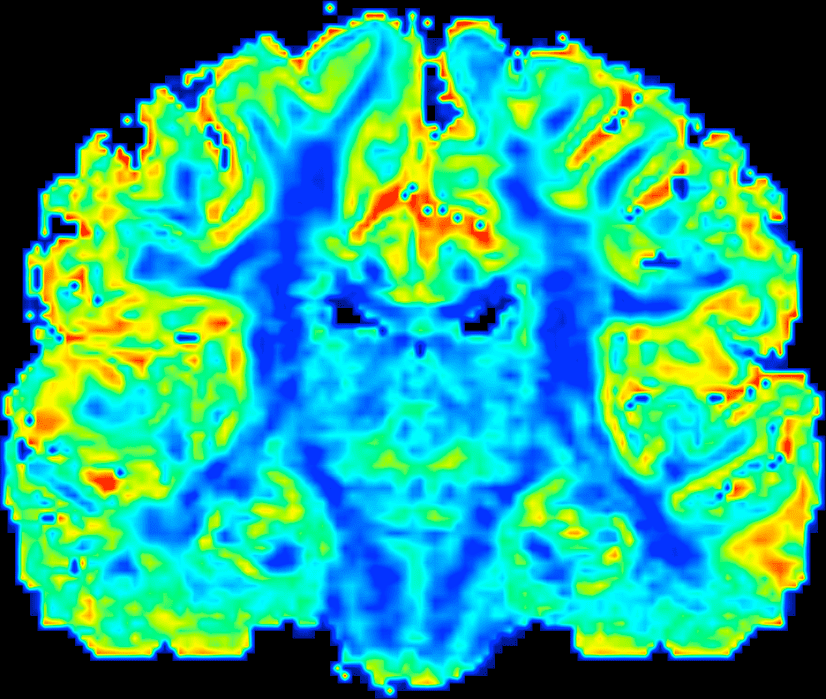}\\[1ex]
			\includegraphics[width=0.8\textwidth]{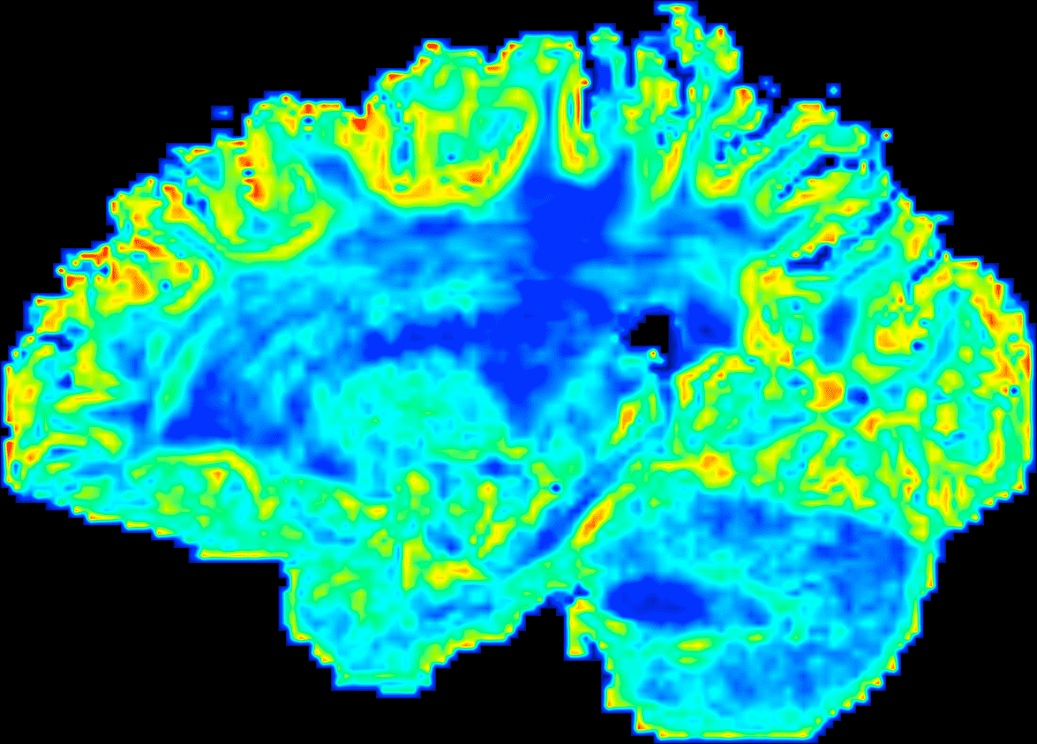}
		\end{minipage}&
		\begin{minipage}{0.3\textwidth}
			\includegraphics[width=0.7\textwidth]{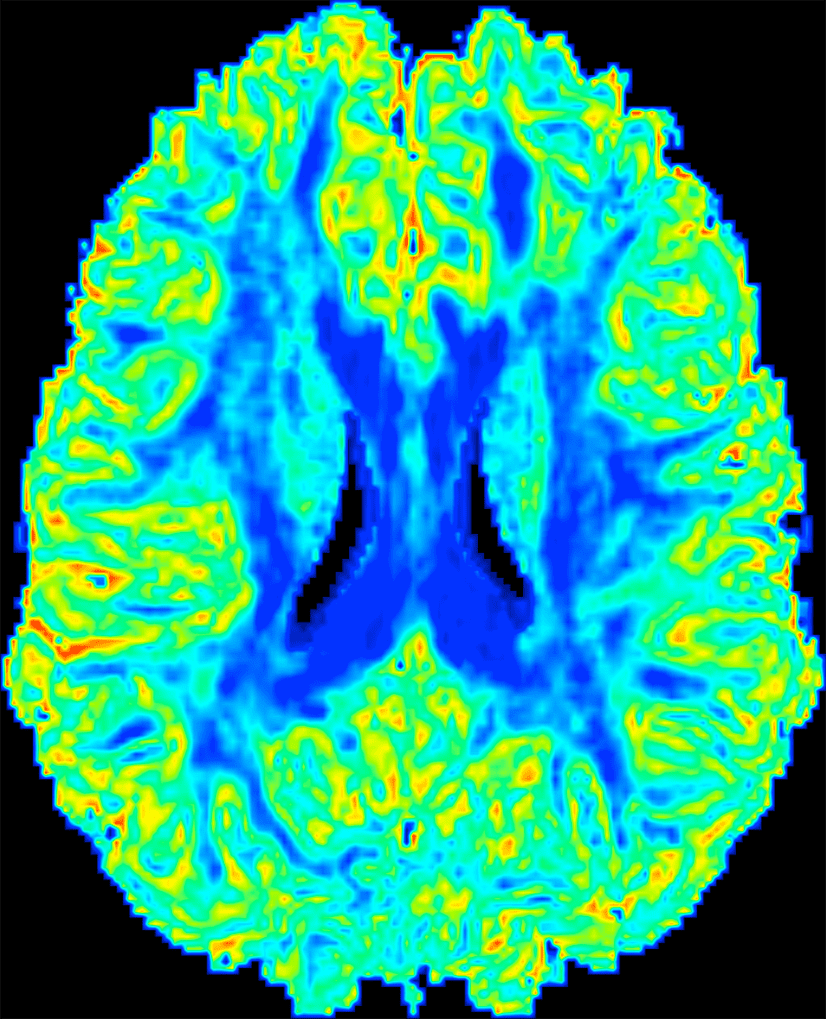}
		\end{minipage}&
		\begin{minipage}{0.2\textwidth}
			\includegraphics[width=0.8\textwidth]{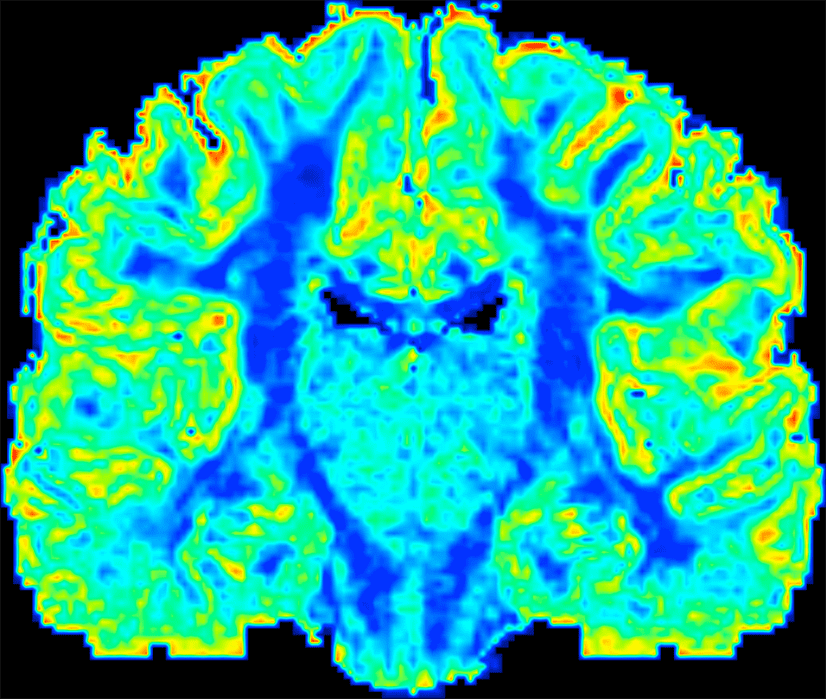}\\[1ex]
			\includegraphics[width=0.8\textwidth]{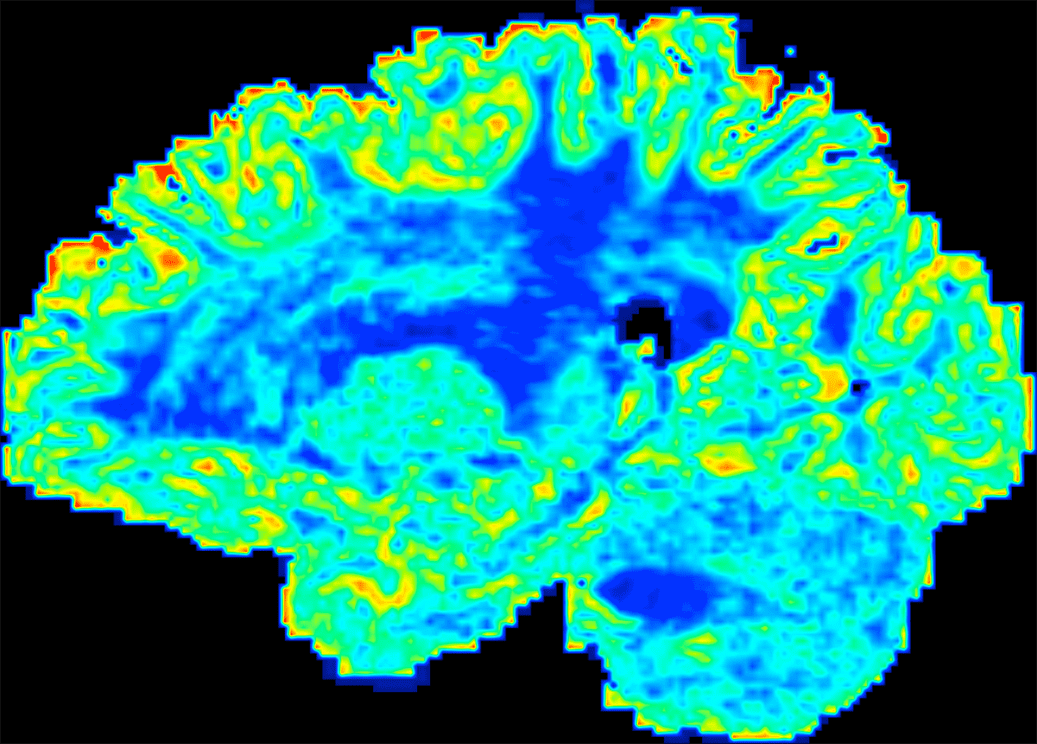}
		\end{minipage}
		\\
		\\[-5pt]
		\hline
		\\[-5pt]
		\rotatebox{90}{\centerline{\textbf{MDI}}} &
		\begin{minipage}{0.3\textwidth}
			\includegraphics[width=0.7\textwidth]{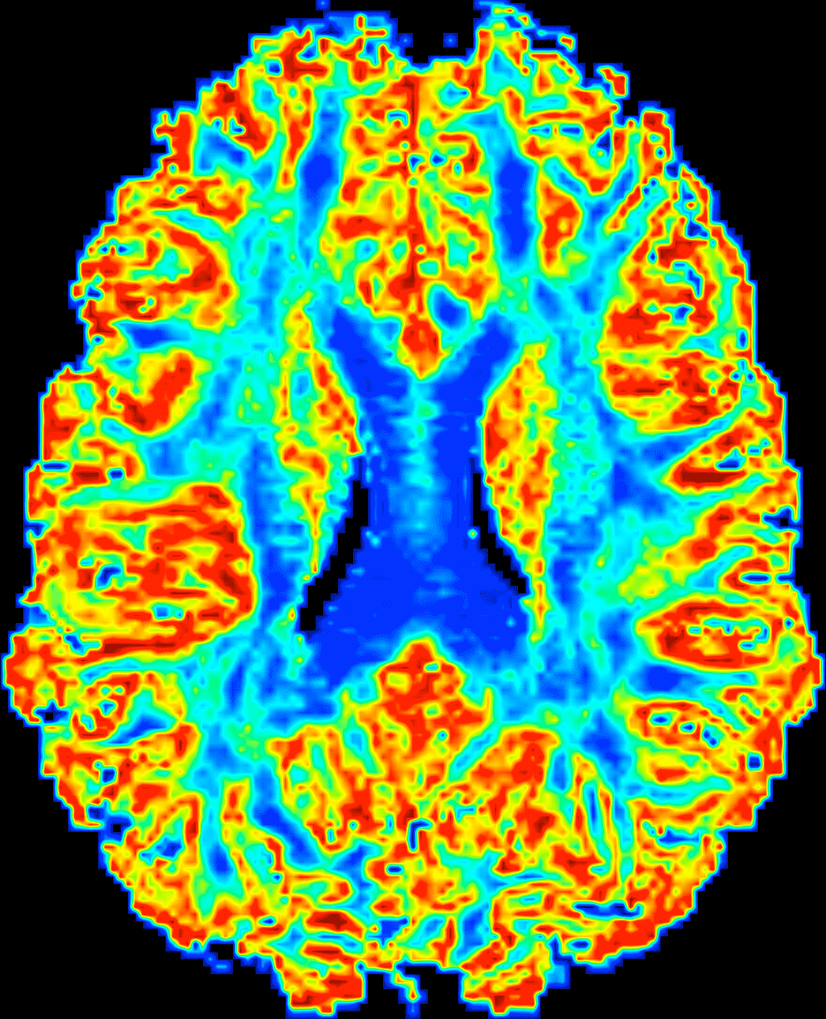}
		\end{minipage}&
		\begin{minipage}{0.2\textwidth}
			\includegraphics[width=0.8\textwidth]{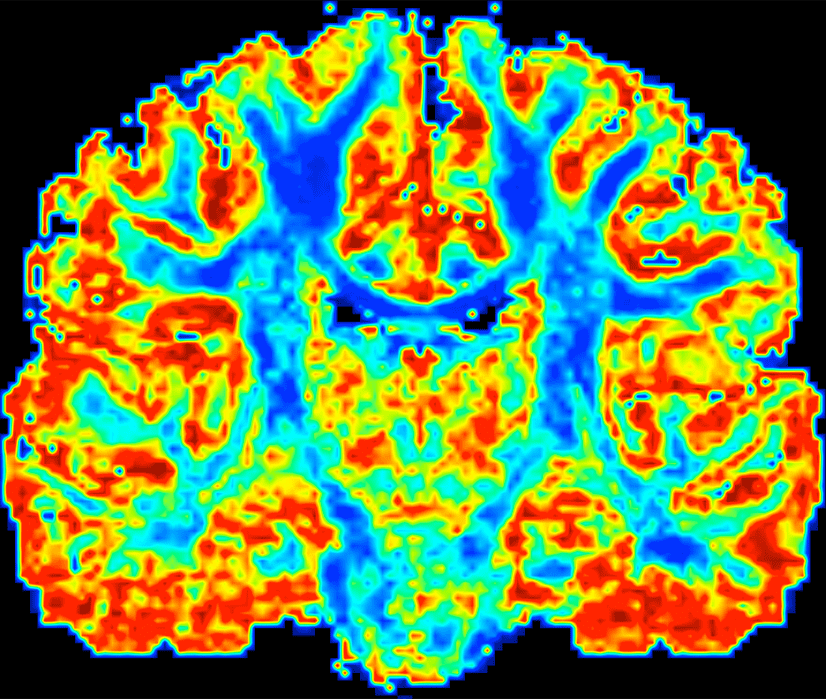}\\[1ex]
			\includegraphics[width=0.8\textwidth]{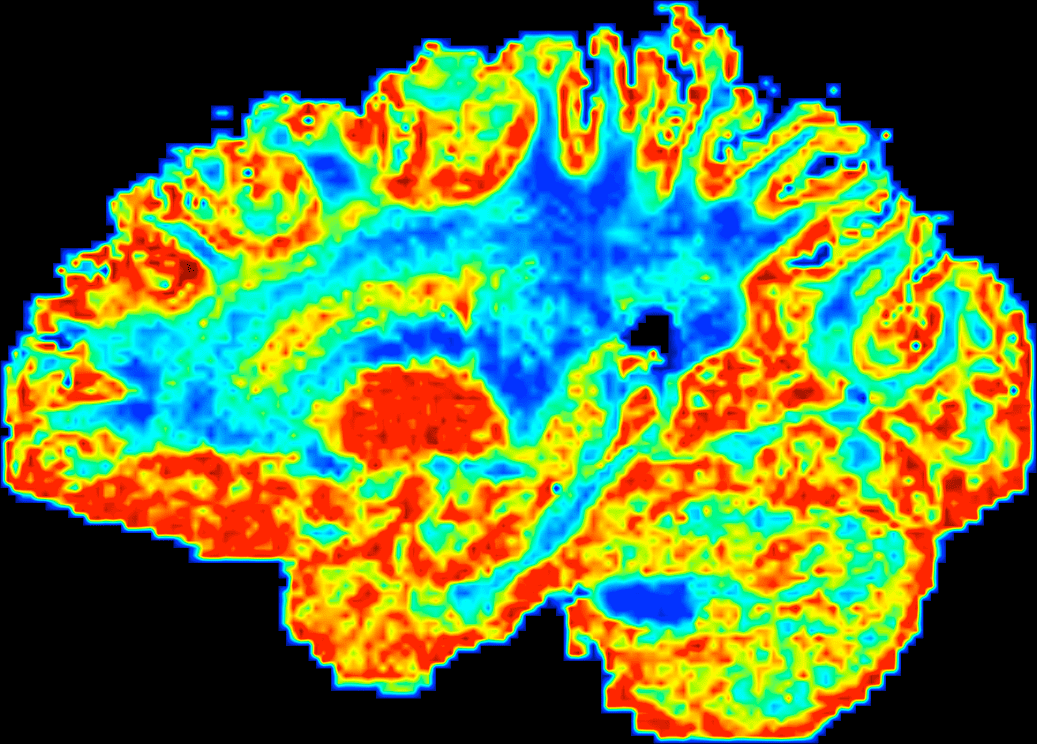}
		\end{minipage}&
		\begin{minipage}{0.3\textwidth}
			\includegraphics[width=0.7\textwidth]{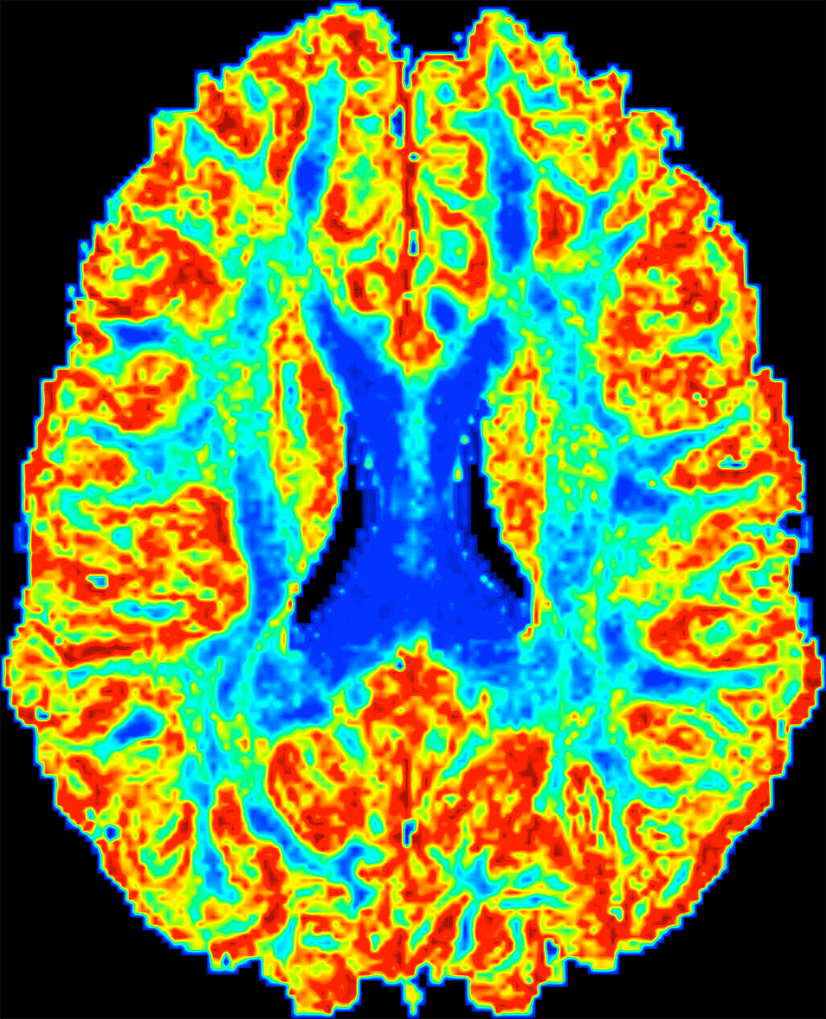}
		\end{minipage}&
		\begin{minipage}{0.2\textwidth}
			\includegraphics[width=0.8\textwidth]{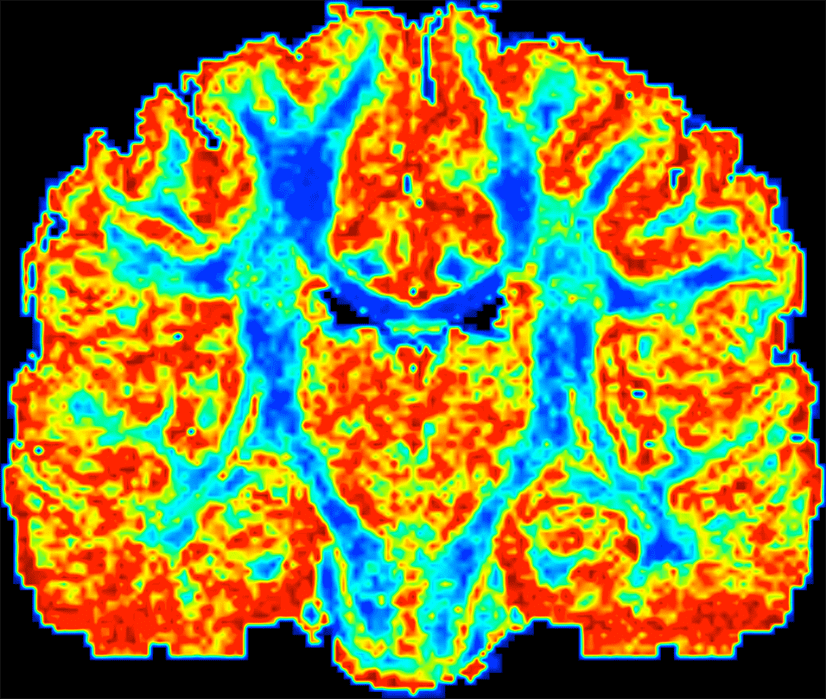}\\[1ex]
			\includegraphics[width=0.8\textwidth]{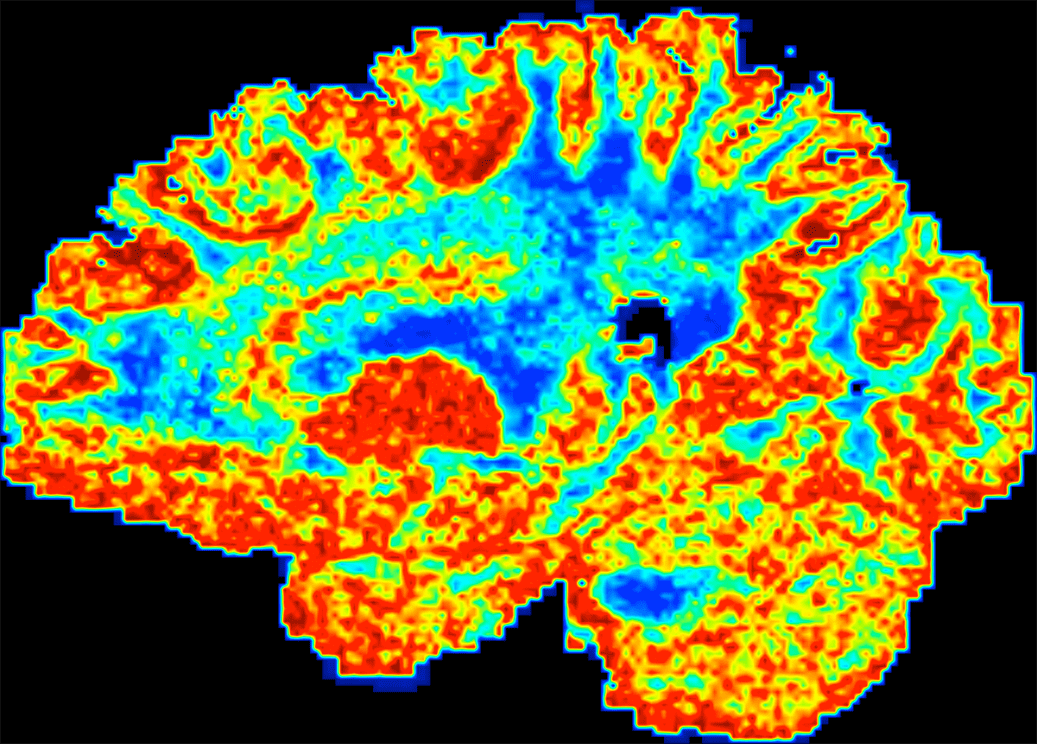}
		\end{minipage}
		\\
		&&&&\colorBar{\tiny0}{\tiny1}{spectrum1}
	\end{tabular}
	
	\caption{FODF asymmetry (ASI) and model discrepancy (MDI) of the 3T and 7T datasets.}\vspace{-10pt}
	
	\label{fig:Indices}
	
\end{figure*}

Figure~\ref{fig:ODF_largeROI} compares the symmetric and asymmetric FODFs based on the HCP data. Close-up views of gyral blades are shown in Figure~\ref{fig:ODF_SmallROI}. It can be observed that the AFODFs reflect the bending and branching characteristics of cortical WM pathways. The asymmetry of the AFODFs becomes more apparent for voxels nearer to the cortex. This is confirmed by the ASI and MDI maps shown in Figure~\ref{fig:Indices}, indicating higher asymmetry and greater model discrepancy in superficial WM than subcortical WM. The asymmetry stems from the curved nature of the axonal trajectories in these regions.

\subsection{Tractography}
\begin{figure*}[!t]\centering
	\begin{tabular}{m{0.01\textwidth}m{0.2\textwidth}m{0.2\textwidth}|m{0.2\textwidth}m{0.2\textwidth}}
		& \centerline{\textbf{FODF - 3T}} & \centerline{\textbf{FODF - 7T}}& \centerline{\textbf{AFODF - 3T}} & \centerline{\textbf{AFODF - 7T}}\\[-20pt]
		\rotatebox{90}{\centerline{\textbf{Axial}}}
		& \includegraphics[width=0.2\textwidth]{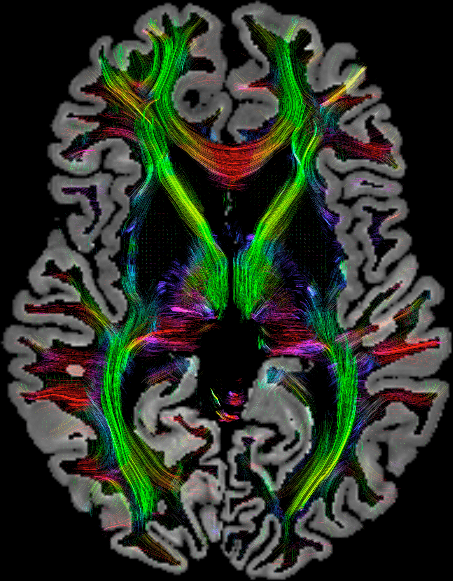} & \includegraphics[width=0.2\textwidth]{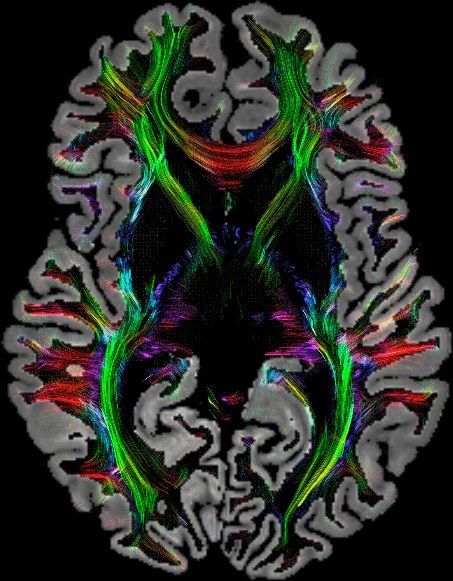}& \includegraphics[width=0.2\textwidth]{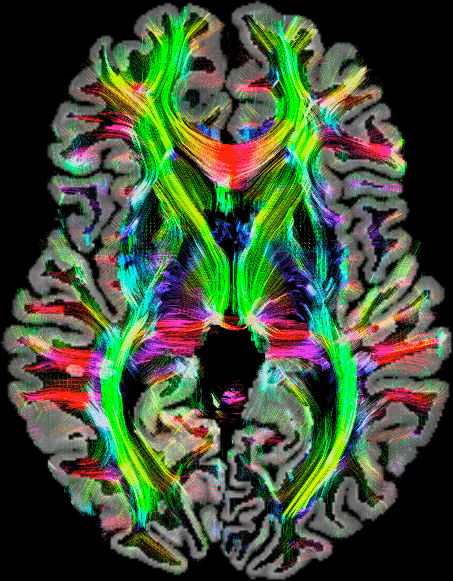} & \includegraphics[width=0.2\textwidth]{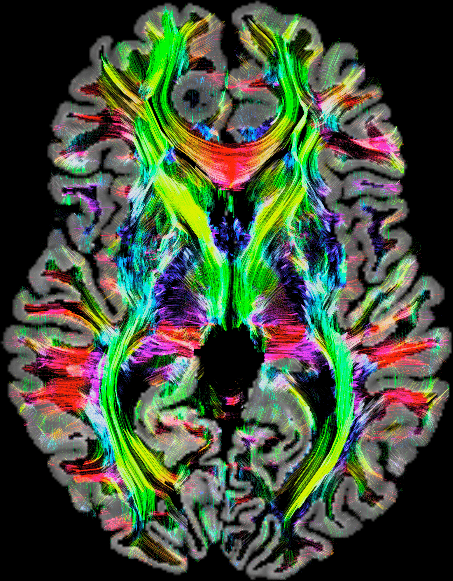}\\
		
		\rotatebox{90}{\centerline{\textbf{Coronal}}}
		& \includegraphics[width=0.2\textwidth]{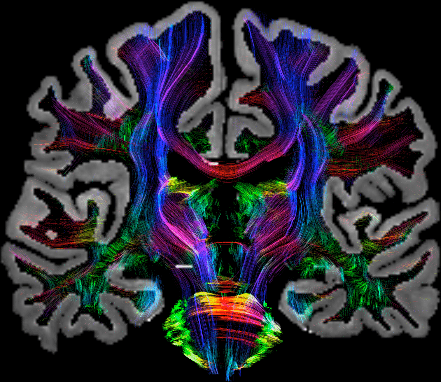} & \includegraphics[width=0.2\textwidth]{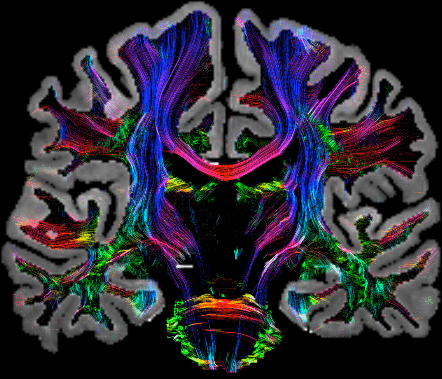}& \includegraphics[width=0.2\textwidth]{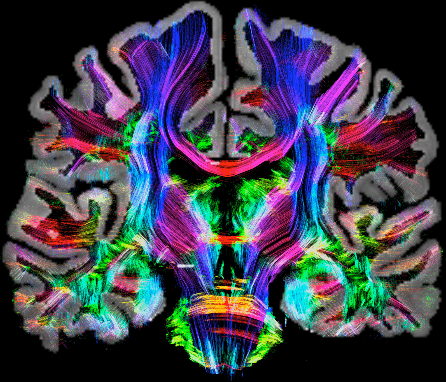} & \includegraphics[width=0.2\textwidth]{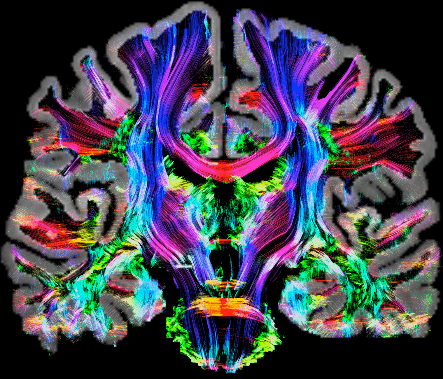}\\
	\end{tabular}
	\caption{Tractography using FODFs and AFODFs. Streamlines within a 5\,mm slab are shown.}
	\label{fig:Tractography}
\end{figure*}

Figure~\ref{fig:Tractography} indicates that
AFODFs yield a significantly greater amount of streamlines. Note that the same GM seeds were used for tractography and we removed streamlines not connecting cortico-cortically and cortico-subcortically or less than 10\,mm. The retained streamlines are denoted as `valid'. The maximum turning angle was set to 50$^{\circ}$, unless mentioned otherwise, for the rest of the paper.
A closer look at the images also shows that symmetric FODFs exhibit gyral bias with streamline endpoints terminating mostly in gyral crowns and much less in sulcal banks.

\begin{figure*}[!th]\centering
	\begin{tabular}{m{0.22\textwidth}m{0.22\textwidth}|m{0.22\textwidth}m{0.22\textwidth}}
		\centerline{\textbf{FODF - 3T}} & \centerline{\textbf{FODF - 7T}} & \centerline{\textbf{AFODF - 3T}} & \centerline{\textbf{AFODF - 7T}} \\ [-20pt]
		\begin{minipage}{0.22\textwidth}
			\includegraphics[width=\textwidth]{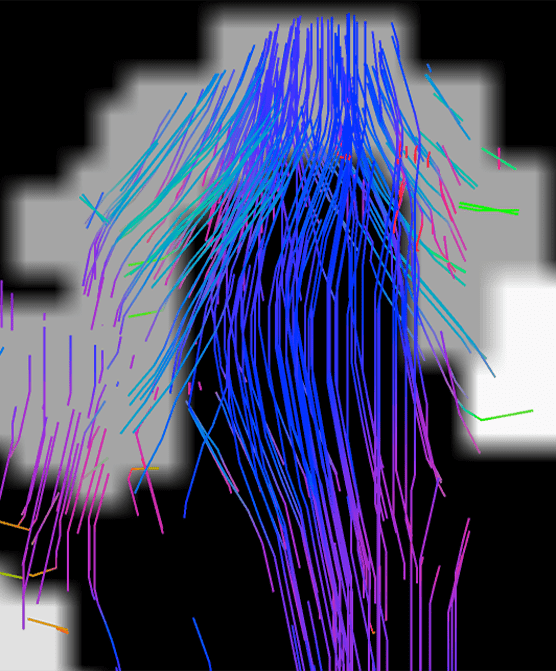}
		\end{minipage}&
		\begin{minipage}{0.22\textwidth}
			\includegraphics[width=\textwidth]{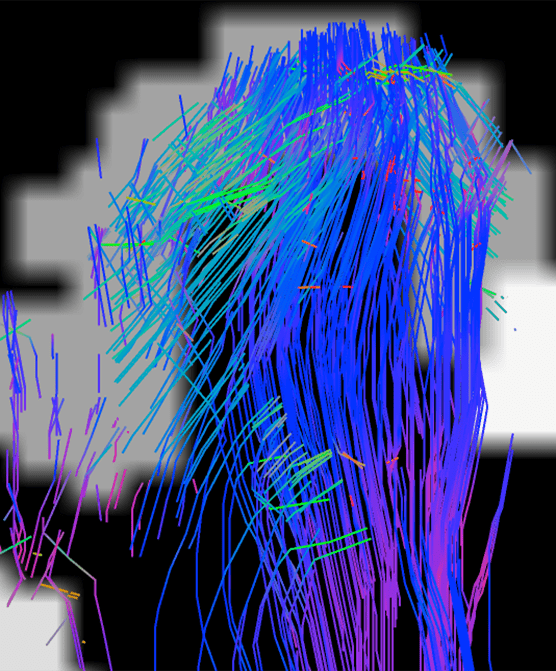}
		\end{minipage}&
		\begin{minipage}{0.22\textwidth}
			\includegraphics[width=\textwidth]{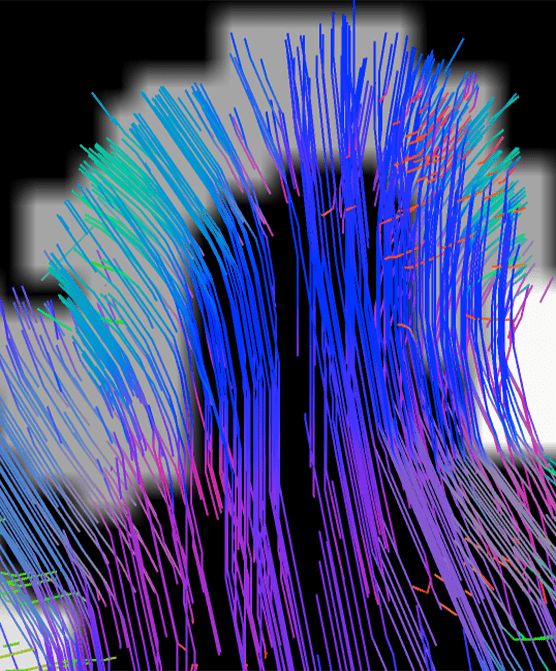}
		\end{minipage}&
		\begin{minipage}{0.22\textwidth}
			\includegraphics[width=\textwidth]{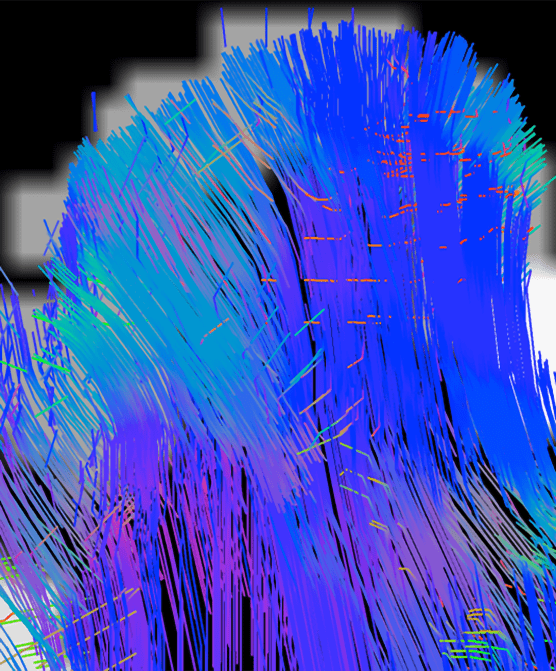}
		\end{minipage}\\
	\end{tabular}
	\caption{Comparison of tractography results in gyral blades. Streamlines within a 5\,mm slab are shown.}
	\label{fig:GyralBias}
\end{figure*}

Figure~\ref{fig:GyralBias} shows tractography results in the gyral blades, indicating that symmetric FODFs are less likely to produce streamlines that turn into the sulcal banks. This problem is mitigated by AFODFs.


\begin{figure*}[!t]\centering
	\begin{tabular}{m{0.22\textwidth}m{0.22\textwidth}|m{0.22\textwidth}m{0.22\textwidth}}
		\centerline{\textbf{FODF - 3T}} & \centerline{\textbf{FODF - 7T}}& \centerline{\textbf{AFODF - 3T}} & \centerline{\textbf{AFODF - 7T}}\\[-20pt]
		\includegraphics[width=0.22\textwidth]{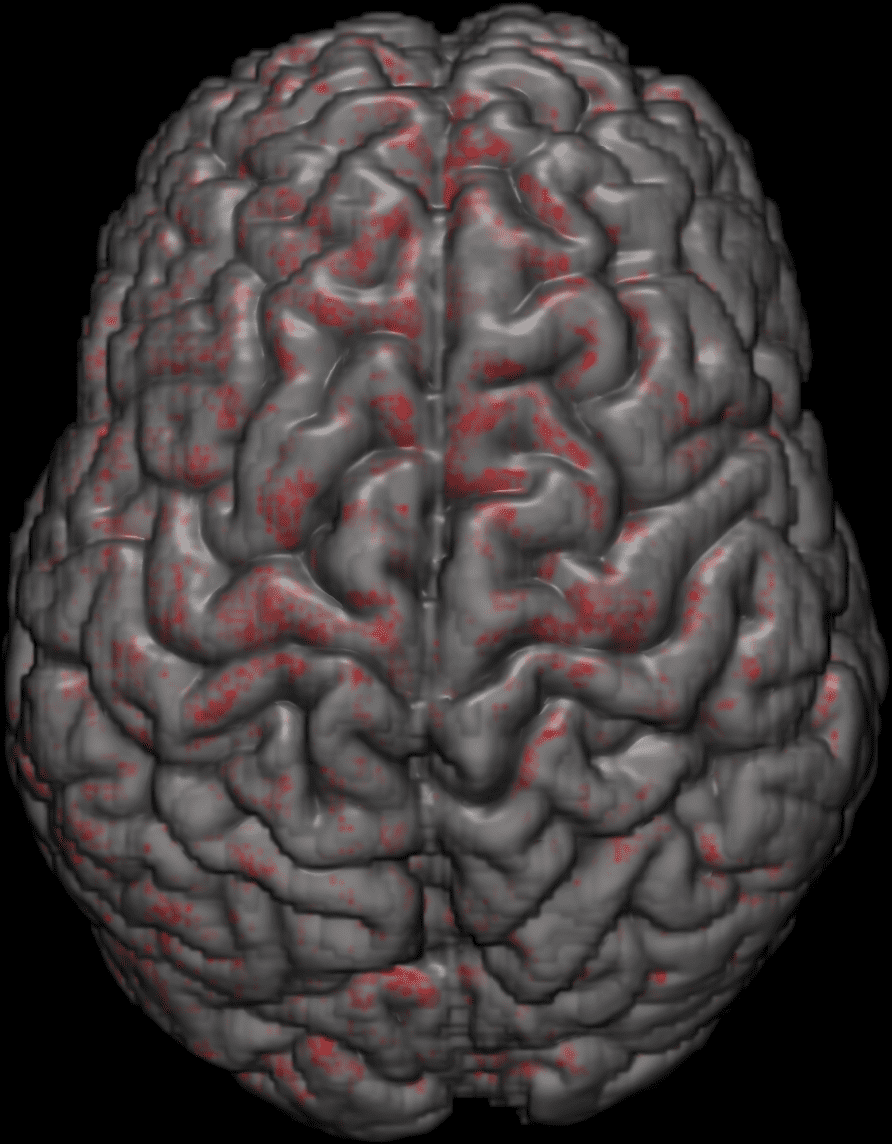} & \includegraphics[width=0.22\textwidth]{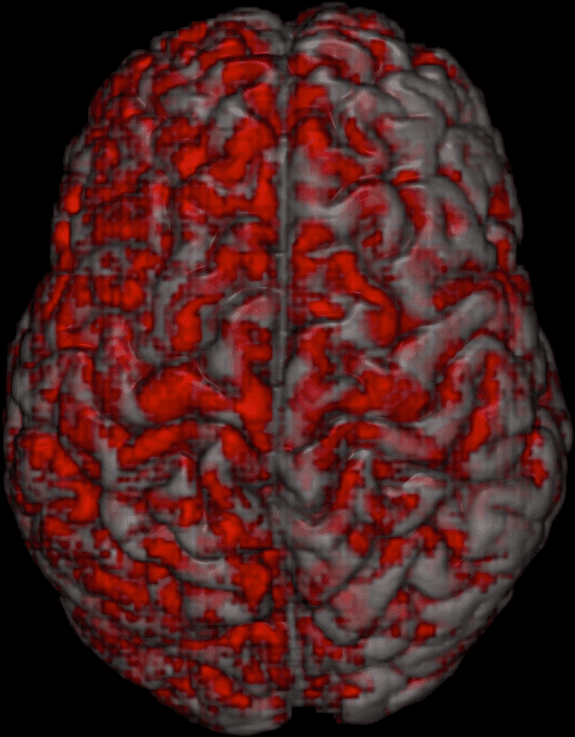}& \includegraphics[width=0.22\textwidth]{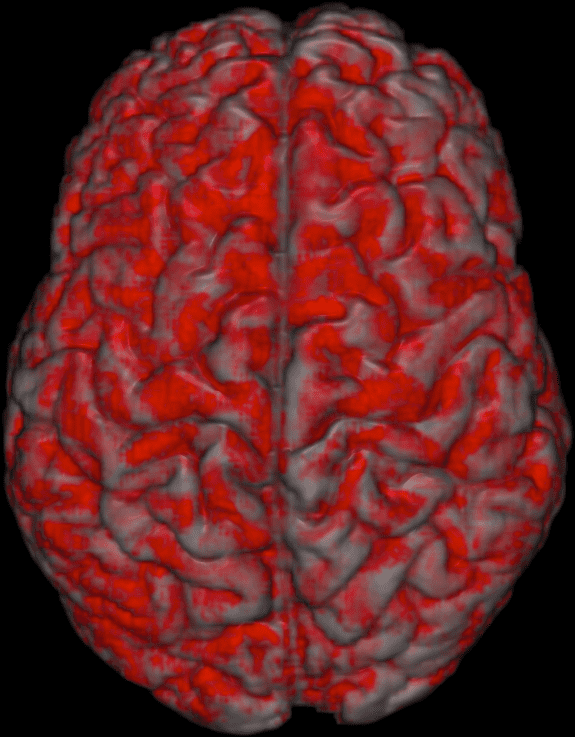} & \includegraphics[width=0.22\textwidth]{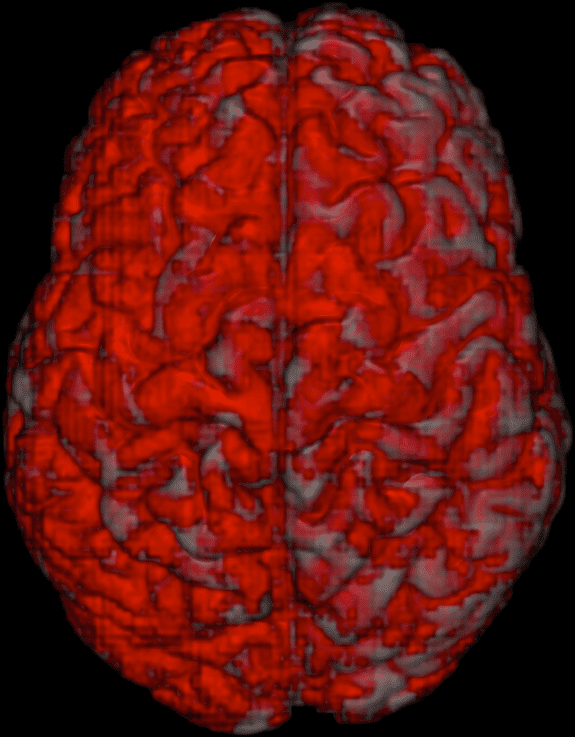}\\
	\end{tabular}
	\caption{Coverage of the cortex by streamline endpoints (red).}
	\label{fig:Coverage}
\end{figure*}

\begin{figure}[!ht]
	\centering
	\pgfplotsset{width=0.7\textwidth,height=0.4\textwidth,compat=1.13}
	\begin{tikzpicture}
	\begin{axis}[
	, ybar
	, xlabel={\textbf{Angle}},
	, ylabel={\textbf{ICI}},
	, ymajorgrids=true
	, axis on top
	, grid style=white
	, major tick length=0pt
	, ymin=0.7, ymax=1
	, xtick={30,50,70}
    , x axis line style={opacity=0}
    , y axis line style={opacity=0}
	, enlarge x limits=0.2
	, xticklabels={30$^\circ$,50$^\circ$,70$^\circ$}
	, legend style={nodes={scale=0.85, transform shape}, at={(0.5, -0.30)}, anchor=north, legend columns=4, cells={anchor=center, fill}}, area legend
	]
	\addplot+[draw opacity=0] table[x = X, y = DET_3T]{plot/rate.txt};\addlegendentry{FODF - 3T}
	\addplot+[draw opacity=0] table[x = X, y = DET_7T]{plot/rate.txt};\addlegendentry{FODF - 7T}
	\addplot+[draw opacity=0] table[x = X, y = aDET_3T]{plot/rate.txt};\addlegendentry{AFODF - 3T}
	\addplot+[draw opacity=0] table[x = X, y = aDET_7T]{plot/rate.txt};\addlegendentry{AFODF - 7T}
	\end{axis}
	\end{tikzpicture}
  \vspace{-5pt}
\caption{Fractions of valid streamlines for various maximum turning angles.}\label{fig:valid_rate}
\end{figure}
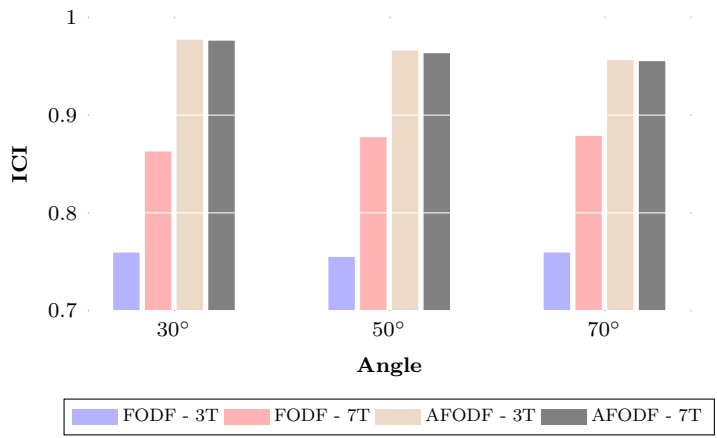

Figure~\ref{fig:Coverage} shows the distribution of the endpoints of fiber streamlines on the cortical surface, confirming that AFODF tractography reduces gyral bias with a more uniform coverage of the cortex across gyral crowns and sulcal banks.
This observation is confirmed quantitatively in Figure~\ref{fig:valid_rate}, which shows that AFODFs yields higher fractions of valid streamlines. This indicates that there is a high probability that a streamline initiated from a GM voxel will connect with another GM voxel.

\subsection{Quantitative Consistency of Endpoints Coverage}
\begin{figure}[!ht]
	\centering
	\pgfplotsset{width=0.7\textwidth,height=0.4\textwidth,compat=1.13}
	
	\begin{tikzpicture}
	\begin{axis}[
	, ybar
	, ylabel={\textbf{ICI}},
	, ymajorgrids=true
	, axis on top
	, grid style=white
	, major tick length=0pt
	, ymin=0.3, ymax=1,
	, xtick={1,2,3}
	, xticklabels={Left Hemisphere,Right Hemisphere,Whole brain}
	, x axis line style={opacity=0}
	, y axis line style={opacity=0}
	, enlarge x limits=0.2
	, legend style={nodes={scale=0.85, transform shape}, at={(0.5, -0.3)}, anchor=north, legend columns=4, cells={anchor=center, fill}}, area legend
	]
	
	\addplot+[draw opacity=0] table[x=X,y=FODF_G]{plot/Consistency_across_scan_noangle.txt};\addlegendentry{FODF - Gyri}
	\addplot+[draw opacity=0] table[x=X,y=FODF_S]{plot/Consistency_across_scan_noangle.txt};\addlegendentry{FODF - Sulci}
	\addplot+[draw opacity=0] table[x=X,y=AFODF_G]{plot/Consistency_across_scan_noangle.txt};\addlegendentry{AFODF - Gyri}
	\addplot+[draw opacity=0] table[x=X,y=AFODF_S]{plot/Consistency_across_scan_noangle.txt};\addlegendentry{AFODF - Sulci}
	
	\end{axis}
	\end{tikzpicture}
	\vspace{-5pt}
	\caption{ICI values for endpoint coverage of gyri and sulci.}\label{fig:consistency_converage}
\end{figure}
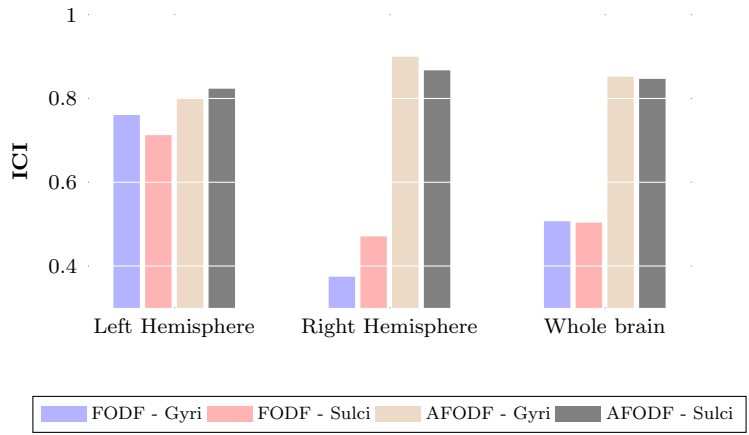

We further investigated the distribution of the streamline endpoints at the cortex.
For the same subject, the distribution should ideally not vary with changes in imaging conditions.
Based on the ROIs given by the Destrieux atlas, we computed the fraction of endpoints in each ROI (i.e., number of endpoints in ROI divided by total number of endpoints over all ROIs). We then computed the ICI based on \eqref{eq:ICI} of the endpoint count fractions of streamlines obtained with symmetric and asymmetric FODFs using datasets acquired with 3T and 7T scanners.
Figure~\ref{fig:consistency_converage} shows the ICI values of streamline endpoints in gyri and sulci, indicating that AFODFs result in greater consistency across field strengths for both gyri and sulci.

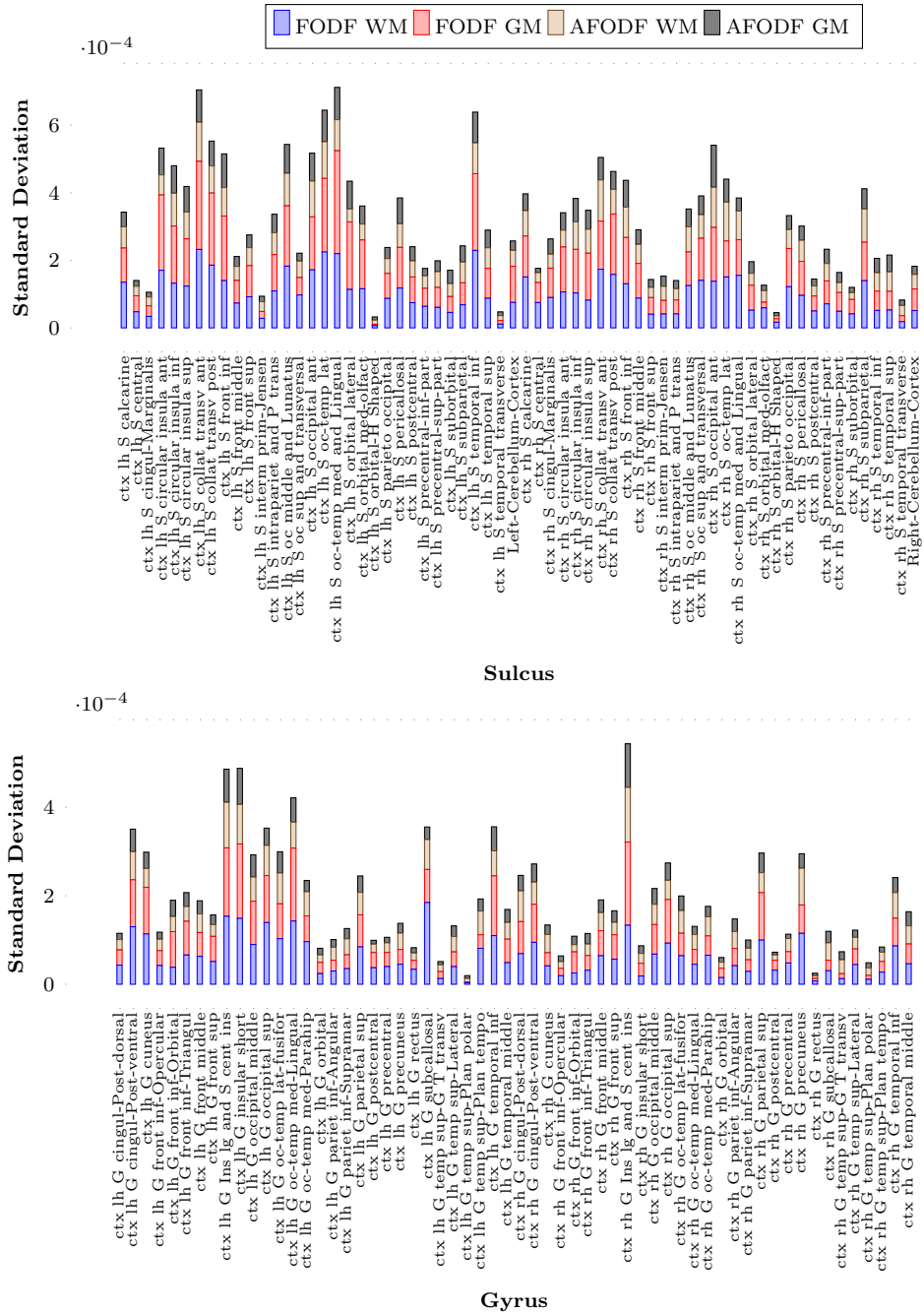
\begin{figure}[t]
	\centering
	\pgfplotsset{width=\textwidth,height=0.4\textwidth,compat=1.13}
	\begin{tikzpicture}
	\begin{axis}[
	ybar stacked,
	grid,
	bar width=2pt,
	legend style={at={(0.55,1.05)},
		anchor=south,legend columns=-1},
	xlabel={\textbf{Sulcus}},
	ylabel={\textbf{Standard Deviation}},
	grid style=white,
	major tick length=0pt,
    enlarge x limits=0.05,
    enlarge y limits=0.1,
    x axis line style={opacity=0},
	y axis line style={opacity=0},
	ymin=0,
	xmin=0,xmax=65,
	xtick=data,
	xticklabels={ctx lh S calcarine,ctx lh S central,ctx lh S cingul-Marginalis,ctx lh S circular insula ant,ctx lh S circular insula inf,ctx lh S circular insula sup,ctx lh S collat transv ant,ctx lh S collat transv post,ctx lh S front inf,ctx lh S front middle,ctx lh S front sup,ctx lh S interm prim-Jensen,ctx lh S intrapariet and P trans,ctx lh S oc middle and Lunatus,ctx lh S oc sup and transversal,ctx lh S occipital ant,ctx lh S oc-temp lat,ctx lh S oc-temp med and Lingual,ctx lh S orbital lateral,ctx lh S orbital med-olfact,ctx lh S orbital-H Shaped,ctx lh S parieto occipital,ctx lh S pericallosal,ctx lh S postcentral,ctx lh S precentral-inf-part,ctx lh S precentral-sup-part,ctx lh S suborbital,ctx lh S subparietal,ctx lh S temporal inf,ctx lh S temporal sup,ctx lh S temporal transverse,Left-Cerebellum-Cortex,ctx rh S calcarine,ctx rh S central,ctx rh S cingul-Marginalis,ctx rh S circular insula ant,ctx rh S circular insula inf,ctx rh S circular insula sup,ctx rh S collat transv ant,ctx rh S collat transv post,ctx rh S front inf,ctx rh S front middle,ctx rh S front sup,ctx rh S interm prim-Jensen,ctx rh S intrapariet and P trans,ctx rh S oc middle and Lunatus,ctx rh S oc sup and transversal,ctx rh S occipital ant,ctx rh S oc-temp lat,ctx rh S oc-temp med and Lingual,ctx rh S orbital lateral,ctx rh S orbital med-olfact,ctx rh S orbital-H Shaped,ctx rh S parieto occipital,ctx rh S pericallosal,ctx rh S postcentral,ctx rh S precentral-inf-part,ctx rh S precentral-sup-part,ctx rh S suborbital,ctx rh S subparietal,ctx rh S temporal inf,ctx rh S temporal sup,ctx rh S temporal transverse,Right-Cerebellum-Cortex},
	xticklabel style = {font=\tiny,yshift=0.5ex},
	x tick label style={rotate=90,anchor=east},
	]
	\addplot+[ybar] table[x=X,y=FODF_WM]{plot/HCP_FD_Sulcal.txt};\addlegendentry{FODF WM}
	\addplot+[ybar] table[x=X,y=FODF_GM]{plot/HCP_FD_Sulcal.txt};\addlegendentry{FODF GM}
	\addplot+[ybar] table[x=X,y=AFODF_WM]{plot/HCP_FD_Sulcal.txt};\addlegendentry{AFODF WM}
	\addplot+[ybar] table[x=X,y=AFODF_GM]{plot/HCP_FD_Sulcal.txt};\addlegendentry{AFODF GM}
	\end{axis}
	\end{tikzpicture}

	\begin{tikzpicture}
	\begin{axis}[
	ybar stacked,
	grid,
	bar width=2pt,
	legend style={at={(0.55,1.05)},
		anchor=south,legend columns=-1},
	xlabel={\textbf{Gyrus}},
	ylabel={\textbf{Standard Deviation}},
	grid style=white,
	major tick length=0pt,
	enlarge x limits=0.05,
	enlarge y limits=0.1,
	x axis line style={opacity=0},
	y axis line style={opacity=0},
	ymin=0,
	xmin=0,xmax=61,
	xtick=data,
	xticklabels={ctx lh G cingul-Post-dorsal,ctx lh G cingul-Post-ventral,ctx lh G cuneus,ctx lh G front inf-Opercular,ctx lh G front inf-Orbital,ctx lh G front inf-Triangul,ctx lh G front middle,ctx lh G front sup,ctx lh G Ins lg and S cent ins,ctx lh G insular short,ctx lh G occipital middle,ctx lh G occipital sup,ctx lh G oc-temp lat-fusifor,ctx lh G oc-temp med-Lingual,ctx lh G oc-temp med-Parahip,ctx lh G orbital,ctx lh G pariet inf-Angular,ctx lh G pariet inf-Supramar,ctx lh G parietal sup,ctx lh G postcentral,ctx lh G precentral,ctx lh G precuneus,ctx lh G rectus,ctx lh G subcallosal,ctx lh G temp sup-G T transv,ctx lh G temp sup-Lateral,ctx lh G temp sup-Plan polar,ctx lh G temp sup-Plan tempo,ctx lh G temporal inf,ctx lh G temporal middle,ctx rh G cingul-Post-dorsal,ctx rh G cingul-Post-ventral,ctx rh G cuneus,ctx rh G front inf-Opercular,ctx rh G front inf-Orbital,ctx rh G front inf-Triangul,ctx rh G front middle,ctx rh G front sup,ctx rh G Ins lg and S cent ins,ctx rh G insular short,ctx rh G occipital middle,ctx rh G occipital sup,ctx rh G oc-temp lat-fusifor,ctx rh G oc-temp med-Lingual,ctx rh G oc-temp med-Parahip,ctx rh G orbital,ctx rh G pariet inf-Angular,ctx rh G pariet inf-Supramar,ctx rh G parietal sup,ctx rh G postcentral,ctx rh G precentral,ctx rh G precuneus,ctx rh G rectus,ctx rh G subcallosal,ctx rh G temp sup-G T transv,ctx rh G temp sup-Lateral,ctx rh G temp sup-Plan polar,ctx rh G temp sup-Plan tempo,ctx rh G temporal inf,ctx rh G temporal middle},
	xticklabel style = {font=\tiny,yshift=0.5ex},
	x tick label style={rotate=90,anchor=east},
	]
	\addplot+[ybar] table[x=X,y=FODF_WM]{plot/HCP_FD_Gyral.txt};
	\addplot+[ybar] table[x=X,y=FODF_GM]{plot/HCP_FD_Gyral.txt};
	\addplot+[ybar] table[x=X,y=AFODF_WM]{plot/HCP_FD_Gyral.txt};
	\addplot+[ybar] table[x=X,y=AFODF_GM]{plot/HCP_FD_Gyral.txt};
	\end{axis}
	\end{tikzpicture}
  \vspace{-5pt}
\caption{Standard deviations of fiber densities at gyral and sulcul WM-GM boundaries.}\label{fig:hcp_fd}
\end{figure}

\begin{figure}[t]
	\centering
	\pgfplotsset{width=\textwidth,height=0.4\textwidth,compat=1.13}
	\begin{tikzpicture}
	\begin{axis}[
	ybar stacked,
	grid,
	bar width=2pt,
	legend style={at={(0.55,1.05)},
		anchor=south,legend columns=-1},
	xlabel={\textbf{Sulcus}},
	ylabel={\textbf{Mean Density}},
	grid style=white,
	major tick length=0pt,
    enlarge x limits=0.05,
    enlarge y limits=0.1,
    x axis line style={opacity=0},
	y axis line style={opacity=0},
	ymin=0,
	xmin=0,xmax=65,
	xtick=data,
	xticklabels={ctx lh S calcarine,ctx lh S central,ctx lh S cingul-Marginalis,ctx lh S circular insula ant,ctx lh S circular insula inf,ctx lh S circular insula sup,ctx lh S collat transv ant,ctx lh S collat transv post,ctx lh S front inf,ctx lh S front middle,ctx lh S front sup,ctx lh S interm prim-Jensen,ctx lh S intrapariet and P trans,ctx lh S oc middle and Lunatus,ctx lh S oc sup and transversal,ctx lh S occipital ant,ctx lh S oc-temp lat,ctx lh S oc-temp med and Lingual,ctx lh S orbital lateral,ctx lh S orbital med-olfact,ctx lh S orbital-H Shaped,ctx lh S parieto occipital,ctx lh S pericallosal,ctx lh S postcentral,ctx lh S precentral-inf-part,ctx lh S precentral-sup-part,ctx lh S suborbital,ctx lh S subparietal,ctx lh S temporal inf,ctx lh S temporal sup,ctx lh S temporal transverse,Left-Cerebellum-Cortex,ctx rh S calcarine,ctx rh S central,ctx rh S cingul-Marginalis,ctx rh S circulafr insula ant,ctx rh S circular insula inf,ctx rh S circular insula sup,ctx rh S collat transv ant,ctx rh S collat transv post,ctx rh S front inf,ctx rh S front middle,ctx rh S front sup,ctx rh S interm prim-Jensen,ctx rh S intrapariet and P trans,ctx rh S oc middle and Lunatus,ctx rh S oc sup and transversal,ctx rh S occipital ant,ctx rh S oc-temp lat,ctx rh S oc-temp med and Lingual,ctx rh S orbital lateral,ctx rh S orbital med-olfact,ctx rh S orbital-H Shaped,ctx rh S parieto occipital,ctx rh S pericallosal,ctx rh S postcentral,ctx rh S precentral-inf-part,ctx rh S precentral-sup-part,ctx rh S suborbital,ctx rh S subparietal,ctx rh S temporal inf,ctx rh S temporal sup,ctx rh S temporal transverse,Right-Cerebellum-Cortex},
	xticklabel style = {font=\tiny,yshift=0.5ex},
	x tick label style={rotate=90,anchor=east},
	]
	\addplot+[ybar] table[x=X,y=FODF_WM]{plot/HCP_MD_Sulcal.txt};\addlegendentry{FODF WM}
	\addplot+[ybar] table[x=X,y=FODF_GM]{plot/HCP_MD_Sulcal.txt};\addlegendentry{FODF GM}
	\addplot+[ybar] table[x=X,y=AFODF_WM]{plot/HCP_MD_Sulcal.txt};\addlegendentry{AFODF WM}
	\addplot+[ybar] table[x=X,y=AFODF_GM]{plot/HCP_MD_Sulcal.txt};\addlegendentry{AFODF GM}
	\end{axis}
	\end{tikzpicture}

	\begin{tikzpicture}
	\begin{axis}[
	ybar stacked,
	grid,
	bar width=2pt,
	legend style={at={(0.55,1.05)},
		anchor=south,legend columns=-1},
	xlabel={\textbf{Gyrus}},
	ylabel={\textbf{Mean Density}},
	grid style=white,
	major tick length=0pt,
	enlarge x limits=0.05,
	enlarge y limits=0.1,
	x axis line style={opacity=0},
	y axis line style={opacity=0},
	ymin=0,
	xmin=0,xmax=61,
	xtick=data,
	xticklabels={ctx lh G cingul-Post-dorsal,ctx lh G cingul-Post-ventral,ctx lh G cuneus,ctx lh G front inf-Opercular,ctx lh G front inf-Orbital,ctx lh G front inf-Triangul,ctx lh G front middle,ctx lh G front sup,ctx lh G Ins lg and S cent ins,ctx lh G insular short,ctx lh G occipital middle,ctx lh G occipital sup,ctx lh G oc-temp lat-fusifor,ctx lh G oc-temp med-Lingual,ctx lh G oc-temp med-Parahip,ctx lh G orbital,ctx lh G pariet inf-Angular,ctx lh G pariet inf-Supramar,ctx lh G parietal sup,ctx lh G postcentral,ctx lh G precentral,ctx lh G precuneus,ctx lh G rectus,ctx lh G subcallosal,ctx lh G temp sup-G T transv,ctx lh G temp sup-Lateral,ctx lh G temp sup-Plan polar,ctx lh G temp sup-Plan tempo,ctx lh G temporal inf,ctx lh G temporal middle,ctx rh G cingul-Post-dorsal,ctx rh G cingul-Post-ventral,ctx rh G cuneus,ctx rh G front inf-Opercular,ctx rh G front inf-Orbital,ctx rh G front inf-Triangul,ctx rh G front middle,ctx rh G front sup,ctx rh G Ins lg and S cent ins,ctx rh G insular short,ctx rh G occipital middle,ctx rh G occipital sup,ctx rh G oc-temp lat-fusifor,ctx rh G oc-temp med-Lingual,ctx rh G oc-temp med-Parahip,ctx rh G orbital,ctx rh G pariet inf-Angular,ctx rh G pariet inf-Supramar,ctx rh G parietal sup,ctx rh G postcentral,ctx rh G precentral,ctx rh G precuneus,ctx rh G rectus,ctx rh G subcallosal,ctx rh G temp sup-G T transv,ctx rh G temp sup-Lateral,ctx rh G temp sup-Plan polar,ctx rh G temp sup-Plan tempo,ctx rh G temporal inf,ctx rh G temporal middle},
	xticklabel style = {font=\tiny,yshift=0.5ex},
	x tick label style={rotate=90,anchor=east},
	]
	\addplot+[ybar] table[x=X,y=FODF_WM]{plot/HCP_MD_Gyral.txt};
	\addplot+[ybar] table[x=X,y=FODF_GM]{plot/HCP_MD_Gyral.txt};
	\addplot+[ybar] table[x=X,y=AFODF_WM]{plot/HCP_MD_Gyral.txt};
	\addplot+[ybar] table[x=X,y=AFODF_GM]{plot/HCP_MD_Gyral.txt};
	\end{axis}
	\end{tikzpicture}
  \vspace{-5pt}
\caption{Means of fiber densities at gyral and sulcul WM-GM boundaries.}\label{fig:hcp_md}
\end{figure}
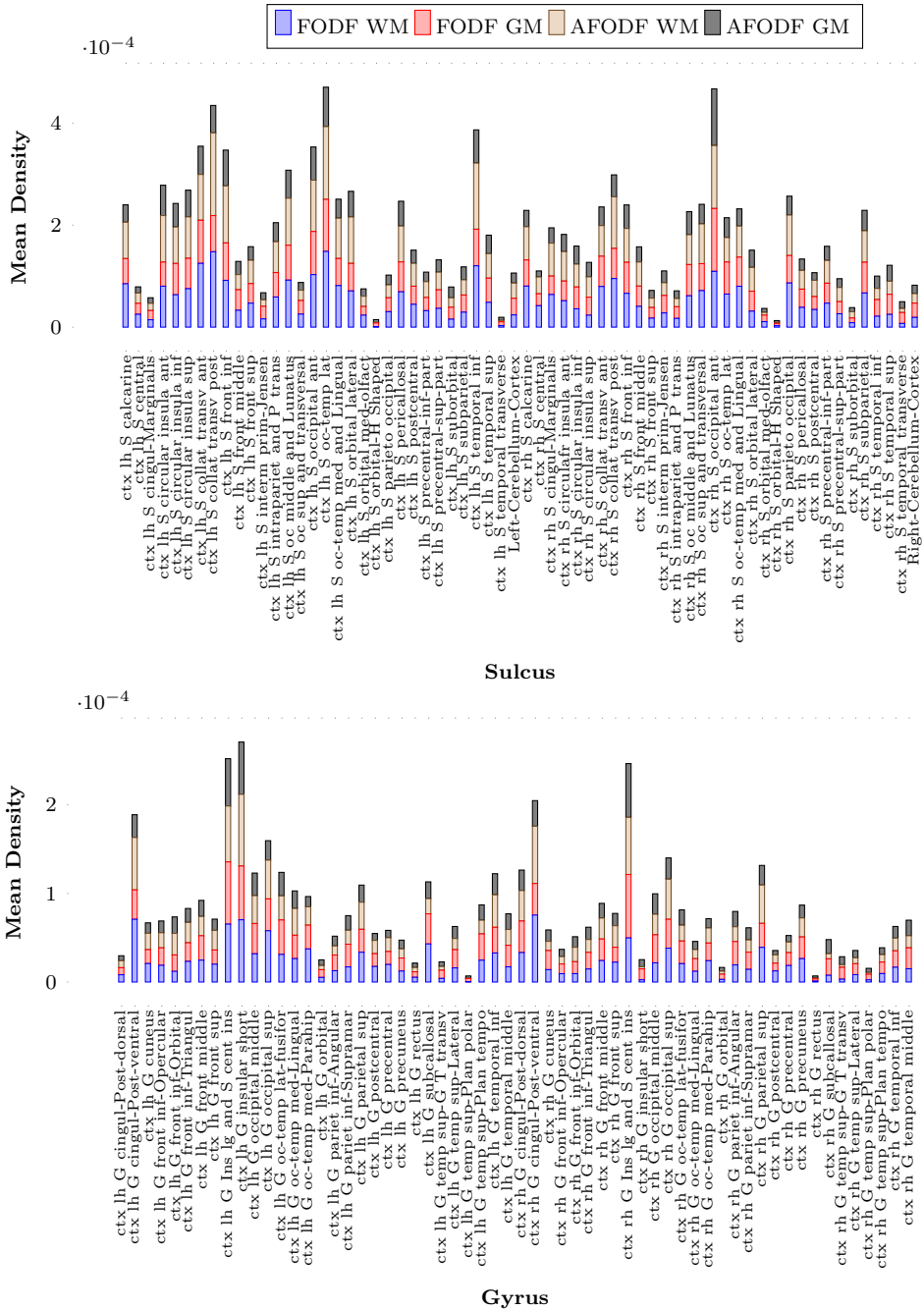

To evaluate the uniformity of the spatial distribution of the streamline endpoints at the cortex, we also computed the standard deviations of the fiber densities (count per volume) across voxels in different cortical regions. GM and WM masks were generated for tractography seeding. The results in Figure~\ref{fig:hcp_fd} indicate that AFODF tractography yields much lower standard deviations than FODF tractography, although these methods show similar means of fiber densities in each region, as shown in Figure~\ref{fig:hcp_md}.


\subsection{Connectivity}
\begin{figure}[t]
\centering
\pgfplotsset{width=0.6\textwidth,height=0.4\textwidth,compat=1.13}
\begin{tikzpicture}
\begin{axis}[
	, ybar
	, xlabel={\textbf{Angle}},
	, ylabel={\textbf{ICI}},
	, ymajorgrids=true
	, axis on top
	, grid style=white
	, major tick length=0pt
	, nodes near coords
	, every node near coord/.append style={rotate=90, anchor=west}
	, ymin=0.8, ymax=1,
    , x axis line style={opacity=0}
	, y axis line style={opacity=0}
	, xtick={30,50,70}
	, xticklabels={30$^\circ$,50$^\circ$,70$^\circ$}
	, legend style={at={(0.5, -0.30)}, anchor=north, legend columns=2, cells={anchor=center, fill}}, area legend
	]
	\addplot+[draw opacity=0] table[x=X,y=FODF_All]{plot/Consistency_connectome.txt};\addlegendentry{FODF}
	\addplot+[draw opacity=0] table[x=X,y=AFODF_All]{plot/Consistency_connectome.txt};\addlegendentry{AFODF}
\end{axis}
\end{tikzpicture}
\vspace{-5pt}
 \caption{ICI values of connectivity matrices.}
\label{fig:ICI_connectivity}
\end{figure}
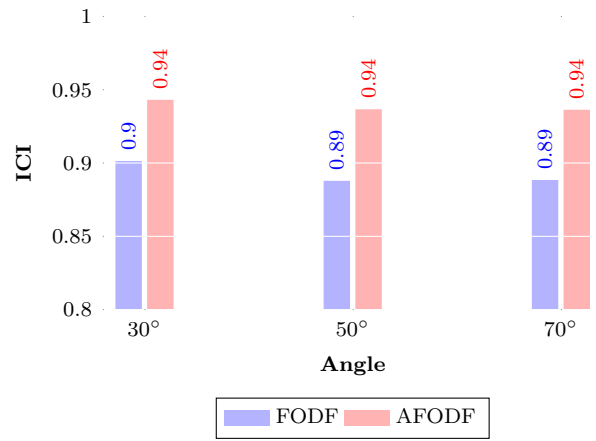

Based on normalized streamlines counts, we constructed a connectivity matrix ($\mathbf{A}$) computed by counting streamlines connecting two regions.
The connectivity matrix was normalized by $\mathbf{D}^{-\frac{1}{2}}\mathbf{A}\mathbf{D}^{-\frac{1}{2}}$ where $\mathbf{D}$ is the degree matrix.
Figure~\ref{fig:ICI_connectivity} shows the ICI values of the vectorized connectivity matrices, indicating that AFODFs result in higher consistency between 3T and 7T.

\section{Discussion}\label{sec:Discussion}
We have introduced a global estimation framework for asymmetric fiber orientation distribution functions (AFODFs). We have shown that AFODFs can resolve subvoxel fiber configurations and mitigate gyral bias in cortical tractography. We have demonstrated that AFODF tractography is reproducible across datasets scanned with different field strengths.
Effective tractography relies on the successful recovery of fiber orientations within voxels of limited spatial resolution.
Estimation of fiber orientation for subvoxel in the gyral blade is challenging because the micro-environment probed by water molecules is much more complex than within the WM.
Our work supports the fact that spatial information across voxels is useful for inferring subvoxel asymmetric orientations.
This is in line with previous work \citep{yap2014fiber,goh2009estimating} in using spatial regularization to improve orientation estimation.

Our method involves a global optimization problem in the sense that the solutions for all voxels are obtained simultaneously. This is an important feature of our method since the fiber continuity constraint causes the solutions of all voxels to be interdependent.
Unlike existing methods of global orientation estimation \citep{schwab2018joint,schwab2018separable,pesce2018fast,auria2015structured},
our formulation is convex and does not rely on initialization based on symmetric FODFs (see \cite{barmpoutis2008extracting,reisert2012geometry,karayumak2018asymmetric} for examples).
This allows the AFODFs to be estimated directly from the diffusion-weighted images.

We have demonstrated that AFODFs are successful in mitigating gyral bias with fuller coverage of both gyral crowns and sulcal banks (Figure~\ref{fig:Coverage}).
Unlike existing methods for mitigating gyral bias \citep{girard2014towards,smith2012anatomically,heidemann2012k,St-Onge2018,teillac2017novel}, the proposed method does not rely on specialized high-resolution imaging techniques and anatomical priors that are derived for example from T1-weighted images. 

Mitigating gyral bias is also important for reliable inference of anatomical pathways connecting cortical regions \citep{schilling2018confirmation}.
With the proposed method, over 95\% of streamlines initiated from cortical GM voxels connect to another cortical GM voxels (Figure~\ref{fig:valid_rate}) with high consistency across different field strengths  (Figure~\ref{fig:ICI_connectivity}). 


\section{Conclusion}\label{sec:Conclusion}
In this work, we have presented a method for estimation of asymmetric fiber orientation distribution functions (AFODFs) with a multi-tissue global framework to mitigate gyral bias in cortical tractography.
Our method allows robust estimation of realistic subvoxel fiber configurations.  We showed that fiber streamlines at gyral blades are able to make sharper turns in gyral blades into the cortical gray matter with AFODFs than conventional FODFs.
We also showed that AFODF tractography results in better cortico-cortical connectivity.
Experimental results also indicate that our method yields greater consistency across different imaging conditions. Comparison with histological data confirms the efficacy our method.

\section{Acknowledgment}\label{sec:Acknowledge}
This work was supported in part by NIH grants (NS093842, EB022880, EB006733, EB009634, AG041721, MH100217, and AA012388) and an NSFC grant (61379020).

\clearpage

\bibliographystyle{elsarticle-harv}
\bibliography{elsarticle}

\end{document}